\documentclass[11pt]{article}

\usepackage[final]{acl}

\usepackage{times}
\usepackage{latexsym}

\usepackage[T1]{fontenc}

\usepackage[utf8]{inputenc}

\usepackage{microtype}

\usepackage{inconsolata}

\usepackage{graphicx}
\usepackage{amsmath}
\usepackage{comment}
\usepackage{caption}
\usepackage{subcaption}
\usepackage{wrapfig}
\usepackage{makecell}

\usepackage{url}            
\usepackage{booktabs}       
\usepackage{amsfonts}       
\usepackage{nicefrac}       
\usepackage{microtype}      
\usepackage{xcolor}         
\usepackage{graphicx}
\usepackage{amsmath}
\usepackage{float}
\usepackage{appendix}

\usepackage{xcolor}
\usepackage{fontawesome5}

\definecolor{mypurple}{RGB}{128,0,128}

\usepackage{hyperref}

\title{emg2speech: synthesizing speech from electromyography using self-supervised speech models}

\author{
  \textbf{Harshavardhana T. Gowda\textsuperscript{}},
  \textbf{Daniel C. Comstock\textsuperscript{}}, and
  \textbf{Lee M. Miller\textsuperscript{}}\\
\textsuperscript{}University of California, Davis
\\\textbf{Correspondence:} \href{tgharshavardhana@gmail.com}{tgharshavardhana@gmail.com}
}

\begin{document}
\maketitle
\begin{abstract}
We present a neuromuscular speech interface that translates electromyographic (EMG) signals recorded from orofacial muscles during speech articulation directly into audio. We find that self-supervised speech (S3) representations are strongly linearly related to the electrical power of muscle activity: a simple linear mapping predicts EMG power from S3 representations with a correlation of {\em r} = 0.85. In addition, EMG power vectors associated with distinct articulatory gestures form structured, separable clusters. Together, these observations suggest that S3 models implicitly encode articulatory mechanisms, as reflected in EMG activity. Leveraging this structure, we map EMG signals into the S3 representation space and synthesize speech, enabling end-to-end EMG-to-speech generation without explicit articulatory modeling or vocoder training. We demonstrate this system with a participant with amyotrophic lateral sclerosis (ALS), converting orofacial EMG recorded while she \emph{silently} articulated speech into audio. 

\textcolor{magenta}{\Large \faGlobe} \href{https://harshavardhanatg.github.io/emg2speech.github.io/}{\textsc{Project Page}}. \textcolor{magenta}{\Large \faGithub} \href{https://github.com/MyoVerse/emg2speech}{\textsc{GitHub}}. \textcolor{magenta}{\Large \faDatabase} \href{https://osf.io/65vbx/files/box}{\textsc{Data}}.

\end{abstract}

\section{Introduction}

Neural and neuromuscular interfaces hold significant promise for augmenting human abilities to interact and communicate with the external world. Brain-computer interfaces (BCIs), such as the speech neuroprostheses described in \citet{wairagkar2025instantaneous, metzger2023high, willett2023high}, have shown that individuals with conditions such as anarthria or amyotrophic lateral sclerosis can regain functional speech through invasive neural recordings. While these invasive approaches are well suited for individuals with severe paralysis or complete loss of articulatory control, their widespread deployment is limited by the need for surgical implantation, high cost, and clinical risk. In contrast, we propose a non-invasive speech interface that leverages preserved articulatory muscle activity, enabling a broader range of individuals, including those with laryngectomy, dysarthria, or dysphonia, to regain functional speech without surgical intervention.

In this article, we present a method that leverages self-supervised speech (S3) models to convert electromyographic (EMG) signals collected during speech articulation directly into audio, without explicitly training a vocoder. Our key insight comes from the observation that speech features derived from S3 models can be linearly mapped to the electrical power of muscle action potentials. Because EMG power patterns associated with different articulatory gestures form structured and separable clusters in feature space, these results suggest that S3 models implicitly encode articulatory information, as reflected in EMG activity. This relationship motivates EMG power as an effective intermediate representation for mapping muscle activity to speech features. We exploit this property to design a lightweight EMG-to-audio conversion model that leverages EMG power representations in conjunction with S3 models.

\section{Prior work}

Converting non-speech signals into audio has been explored across several modalities, including lip-movements-to-speech \citep{kim2021lip, prajwal2020learning}, motor-cortex neural signals-to-speech \citep{wairagkar2025instantaneous, metzger2023high, littlejohn2025streaming}, and EMG-to-speech \citep{gaddy2020digital, gaddy2021improved}. Most existing approaches in these domains \citep{kim2021lip, prajwal2020learning, wairagkar2025instantaneous, gaddy2020digital, gaddy2021improved} assume that the alignment between the input signals (e.g., video or neural activity) and the corresponding audio is known. In contrast, we address a more challenging scenario, similar to \citet{metzger2023high, littlejohn2025streaming}, where the alignment between neural activity (in our case, EMG) and speech is \emph{unknown}. This setting requires the model not only to learn the mapping between EMG activity and audio but also to infer the underlying alignment from an exponential search space.

Work in \citet{littlejohn2025streaming, metzger2023high} addresses this alignment-free setting by training an encoder that takes motor-cortex neural signals as input and maps them to discrete HuBERT units \citep{lakhotia-etal-2021-generative}, which are then passed to a pretrained Tacotron \citep{Tacotron} vocoder following the pipeline in \citet{lakhotia-etal-2021-generative}. We adopt a similar high-level pipeline for EMG-to-speech conversion. However, our approach explicitly leverages the \emph{geometric structure} of EMG signals and their relationship to self-supervised (S3) speech representations to design an encoder grounded in articulatory mechanisms.

Despite this progress, prior work faces several practical limitations. For instance, \citet{littlejohn2025streaming, metzger2023high} use a small-vocabulary corpus containing only 1{,}024 words, and in \citet{littlejohn2025streaming} (where motor-cortex neural activity is converted into speech), each test sentence was presented to the model an average of 6.94 times during training. Moreover, these studies are not fully reproducible due to limited implementation details and the lack of public access to the data used in the experiments. Since non-speech-to-speech conversion typically involves multiple components in an end-to-end pipeline, opaque designs make it difficult to reproduce results and to compare methods fairly. These limitations motivate the need for richer public datasets and reproducible benchmarks for fair evaluation, both of which we provide in this work.

\subsection{Our contributions}

We make three primary contributions.

\textbf{First}, we open-source one of the largest high quality EMG-to-speech datasets to date, comprising a large-vocabulary corpus of approximately 9 hours of EMG speech data with over 6{,}800 unique words from a healthy participant, as well as a small-vocabulary corpus of approximately 1 hour of EMG speech data with roughly 300 unique words from a participant with ALS.
To the best of our knowledge, these datasets constitute one of the largest and most comprehensive publicly available resources for EMG-to-speech conversion.

\textbf{Second}, building on these datasets, we develop encoder architectures grounded in articulatory mechanisms that exhibit interpretability and operate effectively in low-data regimes, including settings with as little as 40 minutes of training data from an ALS participant.
This is particularly important given the practical difficulty of collecting large-scale EMG datasets with current sensing technology.
To support learning under limited supervision, we leverage massively pretrained S3 models and use their representations as a structured target space for EMG-to-speech mapping.
We further establish, for the first time, a quantitative relationship between EMG signals and S3 representations, and exploit this structure to guide encoder design.

\textbf{Third}, we introduce phoneme-guided decoding for EMG-to-speech synthesis and show that incorporating phonetic structure improves the quality of the generated audio. This is because phonemes are defined by articulatory configurations, and EMG signals recorded from multiple orofacial muscle sites capture phonetic structure more faithfully than \textsc{HuBERT} audio units.

Unlike prior EMG-to-speech benchmarks \citep{gaddy2020digital, gaddy2021improved}, our approach does not assume known temporal alignment between EMG and audio during training.
This design is motivated by clinically relevant scenarios in which parallel EMG-audio pairs may be unavailable or unreliable, such as laryngectomy (absence of laryngeal voicing) or ALS (degraded acoustic recordings due to bulbar impairment).
As a result, the model must learn without frame-level EMG-audio correspondence, which substantially increases the difficulty of the learning problem.
Overall, our contributions address fundamental aspects of EMG-to-speech modeling and are simple to implement, yet work well with widely available off-the-shelf pretrained components.

Because our study derives audio from text transcripts rather than using time-aligned EMG-audio pairs, and targets an unaligned EMG-to-speech synthesis setting, there are no existing benchmarks that support direct one-to-one comparisons.
Nevertheless, where possible, we compare against representative baselines from recent EMG interface literature, including spectrogram-based feature pipelines from \textsc{emg2qwerty}~\citep{sivakumaremg2qwerty}, to contextualize performance.
In appendix~\ref{apd:Lit}, we additionally provide broader comparisons to prior brain-computer speech interfaces for context, although these are not intended as direct one-to-one comparisons.


\section{Data}
\label{sec:Data}

We use three datasets in this study: (i) a large, general-corpus vocabulary dataset from a healthy participant, denoted $\textsc{Data}_{\textsc{ general}}$; (ii) a small, limited-corpus vocabulary dataset from a participant with ALS, denoted $\textsc{Data}_{\textsc{ ALS}}$; and (iii) a dataset of discrete orofacial gestures underlying speech articulation collected from 12 healthy participants, denoted $\textsc{Data}_{\textsc{ orofacial gestures}}$. Below, we describe each dataset in detail.

\subsection{$\textsc{Data}_{\textsc{ general}}$}
A healthy participant naturally articulates English sentences while the corresponding EMG signals are recorded at 5000~Hz.
We record EMG from 31 muscle sites on the neck, chin, jaw, cheek, and lips using monopolar surface electrodes (see figure~\ref{fig:electrodePlacement} for electrode placement and appendix~\ref{appendix:exp} for additional details).

We adapt the language corpus from \citet{willett2023high}, who demonstrate a speech brain-computer interface by translating motor-cortex activity into speech.
The dataset comprises an English corpus with approximately 6800 unique words and 9660 sentences.
Sentences vary in length, and the participant articulated at a normal speaking rate, averaging 115 words per minute.
We split the dataset into training, validation, and test sets containing 8500, 760, and 400 sentences, respectively.
Sentences in the test set do not appear in the training or validation sets.

The start and end of each sentence are timestamped using mouse clicks.
When the participant is ready to begin, they click the mouse to display the sentence on the screen and mark the start time.
After articulation is complete, they click again to mark the end time; this second click removes the sentence from the screen.
This procedure allows the participant to articulate at their own pace.

\subsection{$\textsc{Data}_{\textsc{ ALS}}$}
A participant diagnosed with amyotrophic lateral sclerosis (ALS) silently articulates English sentences (with overt articulatory movements but no audible output) while we record the corresponding EMG signals at 5000~Hz using the same electrode layout as in $\textsc{Data}_{\textsc{ general}}$.

We construct a small English corpus comprising approximately 300 unique words and 600 sentences. Sentences vary in length, and the participant articulated at her current comfortable speaking rate, averaging 61 words per minute. We split the dataset into training, validation, and test sets containing 500, 40, and 60 sentences, respectively. Test sentences do not appear in the training or validation sets.

\subsection{$\textsc{Data}_{\textsc{ orofacial gestures}}$}
Twelve participants perform 13 distinct orofacial movements, with 10 repetitions per movement. The set of movements includes \texttt{\footnotesize cheeks: puff out}, \texttt{\footnotesize cheeks: suck in}, \texttt{\footnotesize jaw: drop down}, \texttt{\footnotesize jaw: move backward}, \texttt{\footnotesize jaw: move forward}, \texttt{\footnotesize jaw: move left}, \texttt{\footnotesize jaw: move right}, \texttt{\footnotesize lips: pucker}, \texttt{\footnotesize lips: smile}, \texttt{\footnotesize lips: tuck (as if blotting)}, \texttt{\footnotesize tongue: back of lower teeth}, \texttt{\footnotesize tongue: back of upper teeth}, and \texttt{\footnotesize tongue: roof of the mouth}. These movements are selected to span a broad range of articulatory gestures involved in natural speech production, including lip rounding, jaw positioning, and tongue placement, which are essential for producing different phonemes.

This dataset is recorded using 22 electrodes at a sampling rate of 5000~Hz. Signals are recorded from approximately the same muscle sites as in $\textsc{Data}_{\textsc{ general}}$, except that electrodes on the right side of the neck are not used (middle diagram in figure~\ref{fig:electrodePlacement}). Each gesture is performed for a duration of 1.5~{\em s}.

\section{Methods}
\label{sec:Methods}

\subsection{Electromyography (EMG)}
EMG signals are collected by a set of sensors \(\mathcal{V}\) and represented as functions of time \(t\). 
A sequence of EMG signals \(E\) corresponding to articulated speech, associated with an audio signal \(A\) and phonemic content \(L\), is represented as \(E = \{\mathbf{f}_v(t)\}_{\forall \, v \in \mathcal{V}}\).
Here, \(\mathbf{f}_v(t)\) denotes the EMG signal captured at sensor node \(v\) as a function of time. 
The audio signal \(A\) encodes both phonemic (lexical) content and expressive aspects of speech such as volume, pitch, prosody, and intonation, while \(L\) represents only the phonemic content---a sequence of phonemes. 
For example, the phonemic content \(L\) of the word \textsc{$<$friday$>$} is denoted by the phoneme sequence \textsc{$<$f-r-iy-d-ay$>$}.\\

\textbf{EMG covariance matrices:} for an EMG signal \( E_{\mathcal{V} \times \tau} \) collected from \(\mathcal{V}\) sensor nodes over a duration of \(\tau\) samples, we construct a symmetric positive definite (SPD) covariance matrix \(\mathcal{E}_{\mathcal{V} \times \mathcal{V}} = \epsilon E E^\top\), where \(\epsilon\) is a scaling factor. We denote the diagonal of \(\mathcal{E}\) as \(\mathbb{D}(\mathcal{E})\) and its lower triangular part as \(\lfloor \mathcal{E} \rfloor\). The vector \(\mathbb{D}(\mathcal{E})\) represents the muscle action potential power at each electrode \(v \in \mathcal{V}\) during the interval \(\tau\), while the off-diagonal elements capture the pairwise cross-channel covariance, reflecting the spatial co-activation structure across electrodes. A vectorized representation of \(\mathcal{E}\) is denoted as \(\texttt{vec}(\mathcal{E})\), a column vector of dimension \(\mathcal{V}^2\).

The geodesic distance between two SPD matrices \(\mathcal{E}_1\) and \(\mathcal{E}_2\) is the same as the distance between their corresponding Cholesky matrices \(\mathcal{L}_1\) and \(\mathcal{L}_2\) \citep{lin2019riemannian} and is calculated as 
{
\begin{align}
    d(\mathcal{L}_1, \mathcal{L}_2) &= \left\{||\lfloor \mathcal{L}_1 \rfloor - \lfloor \mathcal{L}_2\rfloor||_F^2 \right. \nonumber \\
    &\quad + \left. ||\log\mathbb{D}(\mathcal{L}_1) - \log\mathbb{D}(\mathcal{L}_2)||_F^2\right\}^{1/2}, \label{eq:distance}
\end{align}
}
where \(||\cdot||_F\) denotes the Frobenius norm. Here, \(\mathcal{L}_1\) and \(\mathcal{L}_2\) are the \emph{Cholesky factors} of the SPD matrices \(\mathcal{E}_1\) and \(\mathcal{E}_2\), i.e., lower triangular matrices such that \(\mathcal{E} = \mathcal{L} \mathcal{L}^\top\).\\

\textbf{EMG spectrograms:} for an EMG signal \( E_{\mathcal{V} \times \tau} \) collected from \(\mathcal{V}\) sensor nodes at a sampling frequency \( f_s \), we compute the short-time Fourier transform (STFT) over successive time windows to obtain a power spectrogram representation \(\mathcal{S}_{\mathcal{V} \times F \times \tau'} = \bigl| \mathrm{STFT}(E_{\mathcal{V} \times \tau}) \bigr|^2\), where \(F\) denotes the number of frequency bins and \(\tau'\) the number of time frames. Each slice \(\mathcal{S}_{\mathcal{V} \times F}^{(t)}\) captures the frequency-domain energy distribution of EMG activity across \(\mathcal{V}\) electrodes at time frame {\em t}. To reduce the spectral granularity, we bin the frequency axis into {\em B} frequency bands to obtain $\mathcal{B}$.  In practice, we use either five bands {\em B}$_{\text{\em 1}}$ = \texttt{[80, 125] Hz}, {\em B}$_{\text{\em 2}}$ = \texttt{[125, 250] Hz}, {\em B}$_{\text{\em 3}}$ = \texttt{[250, 375] Hz}, {\em B}$_{\text{\em 4}}$ = \texttt{[375, 687.5] Hz}, and {\em B}$_{\text{\em 5}}$ = \texttt{[687.5, 1000] Hz}, following \citet{Kaifosh2025Generic}, or 31 linearly spaced frequency bands between \texttt{80 and 1000 Hz}. A vectorized representation of \(\mathcal{B}\) is denoted as \(\texttt{vec}(\mathcal{B})\), a column vector of dimension \(\mathcal{V}\){\em B}.\\

\subsection{Audio (A)}
\textbf{Audio spectrograms:} for a speech waveform \( a(t) \) sampled at frequency \( f_s \), we compute a mel-scaled power spectrogram using a Hann-windowed short-time Fourier transform (STFT), followed by projection onto a mel filterbank with {\em B} mel bands. Specifically, we first obtain the power spectrogram \(\small \mathcal{M}_{F \times \tau'} = \bigl|\mathrm{STFT}(a(t))\bigr|^2\), where {\em F} denotes the number of frequency bins and \( \tau' \) the number of time frames. This spectrogram is then projected onto a mel filterbank \( W_{\mathrm{mel}} \) spanning the frequency range \([f_{\min}, f_{\max}]\), yielding
{
\[
\mathcal{A}_{B \times \tau'}(b, t) = \sum_{f} W_{\mathrm{mel}}(b,f)\,\mathcal{M}_{f,t},
\]}
where \( b \in \{1, \dots, B\} \) indexes the mel bands. Each vector \(\mathcal{A}^{(t)}_{B}\) encodes the mel-band power distribution of the speech signal at frame {\em t}, emphasizing perceptually relevant frequency regions. We use {\em B} = 80 mel bands, \( f_{\min} = 20 \)~Hz, and \( f_{\max} = f_s/2 \). We denote the column vector of an audio spectrogram by \(\mathcal{A}\) throughout the article.\\

\textbf{Audio features from S3 models:} for a speech waveform \( a(t) \), we extract self-supervised speech (S3) representations by passing the signal through a pretrained model \(\mathcal{S}\), yielding \(\mathcal{H} = \mathcal{S}(a(t))\). The model \(\mathcal{S}\) can be instantiated as \textsc{Wav2Vec 2.0} \citep{baevski2020wav2vec}, \textsc{HuBERT} \citep{hsu2021hubert}, or \textsc{WavLM} \citep{chen2022wavlm}. We denote the column vector of S3 audio representations by \(\mathcal{H}\) throughout the article.

\subsection{Sequence-to-sequence models} 
\label{sec:seq2seq}
We construct sequences of \(\texttt{vec}(\mathcal{E})\), \(\texttt{vec}(\mathcal{B})\), \(\mathcal{A}\), and \(\mathcal{H}\), emitted every 20~ms and use a context length of 25~ms. For temporal relation modeling, we employ a causal time depth separable convolutional network (TDS), as described below.

We adapt the TDS model originally designed for EMG-based keyboard typing in~\citet{sivakumaremg2qwerty} with minor modifications. The model relies exclusively on local temporal context, with a 1 {\em s} causal receptive field. To improve robustness to spatial variability in electrode activity, the architecture incorporates a \emph{shift-tolerant} module consisting of a linear layer followed by a ReLU activation. This module is applied to electrode channel shifts of \(-1\), \(0\), and \(+1\) positions, and the resulting outputs are averaged. The concatenated outputs from the shift-tolerant module are then fed into the TDS network for temporal modeling.


\section{Results}
\label{sec:results}

\subsection{\(\mathcal{E}\) and \(\mathbb{D}(\mathcal{E})\) encode articulatory information}
\label{sec:tSNE}

Here we test whether covariance-based EMG features preserve discriminative structure related to articulation. We evaluate this on \(\textsc{Data}_{\textsc{ orofacial gestures}}\), where each trial is an orofacial movement recorded from 22 electrodes over 1.5~{\em s}. Each trial is represented by an EMG signal matrix \(E \in \mathbb{R}^{22 \times 7500}\). We summarize each trial with a symmetric positive definite (SPD) covariance matrix \(\mathcal{E} \in \mathbb{R}^{22 \times 22}\), and additionally consider its diagonal \(\mathbb{D}(\mathcal{E}) \in \mathbb{R}^{22}\), which represents per-channel EMG power.

\begin{figure}[ht] 
    \centering 
    \includegraphics[width=0.42\textwidth]{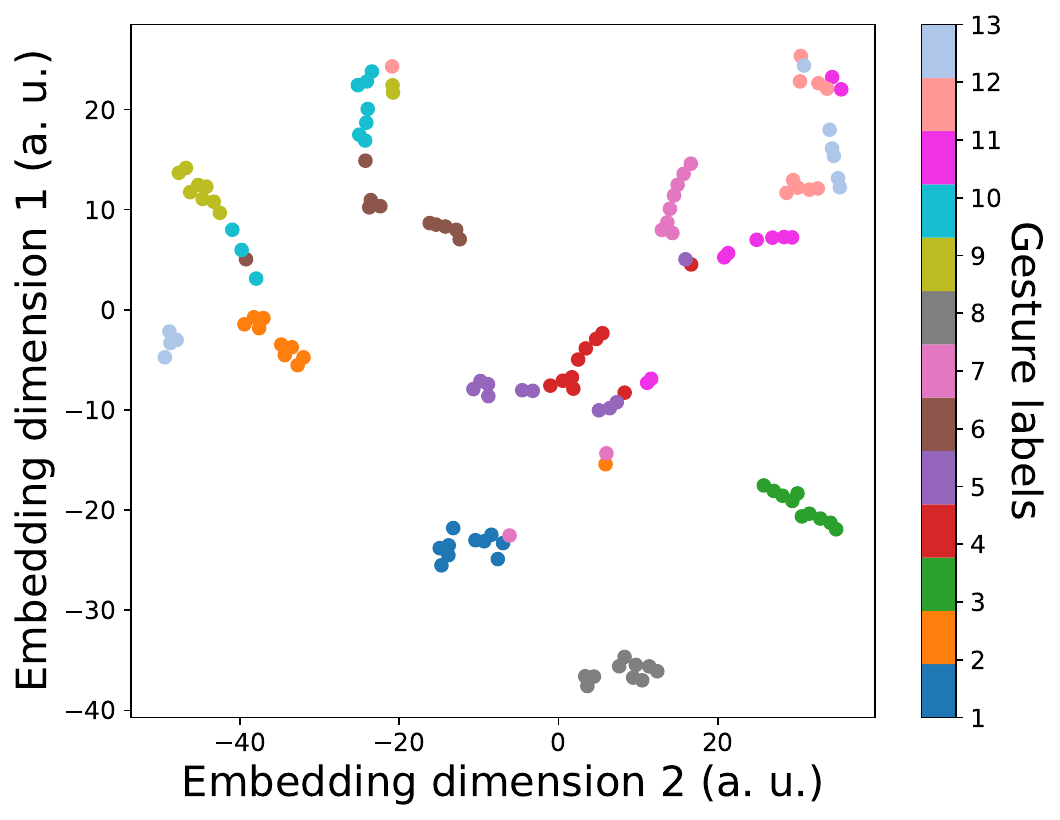} 
    \caption{Different orofacial gestures are naturally separable using covariance-based EMG features. {\em t}-SNE visualization of diagonal vectors \(\mathbb{D}(\mathcal{E})\) for 13 orofacial movements from a single subject. Embedding is color-coded by articulatory gesture type ({\em a.u.} = arbitrary units).}
    \label{fig:oroIndividualDiag} 
\end{figure}

The vectors \(\mathbb{D}(\mathcal{E})\) corresponding to different orofacial gestures naturally form distinct clusters, as shown in figure~\ref{fig:oroIndividualDiag}. We further quantify this separability using the unsupervised {\em k}-medoids clustering algorithm~\citep{Kaufman1990PAM}, achieving an accuracy of 61.41\% using \(\mathbb{D}(\mathcal{E})\) (averaged across 12 subjects). When using the full covariance matrix \(\mathcal{E}\) with the geodesic distance defined in equation~\ref{eq:distance}, the {\em k}-medoids accuracy increases to 73.7\%; both results are well above the random-chance level of 7.69\%.

These results demonstrate that both \(\mathcal{E}\) and \(\mathbb{D}(\mathcal{E})\) encode discriminative articulatory information. While \(\mathbb{D}(\mathcal{E})\) alone is sufficient to distinguish among different orofacial movements, incorporating the full covariance structure in \(\mathcal{E}\) improves decoding accuracy.

Note that other widely used EMG features, such as log-spectrograms~\citep{sivakumaremg2qwerty} or rectified time-domain signals~\citep{Halliday2010Rectification}, are not as straightforward to probe for this type of global structure. When raw EMG signals \(E \in \mathbb{R}^{22 \times 7500}\) are featurized using spectrograms or rectified signals, the temporal dimension may be reduced in granularity but is not collapsed into a single frame. In contrast, covariance-based representations aggregate temporal information into a single fixed-dimensional feature, yielding \(\mathbb{D}(\mathcal{E}) \in \mathbb{R}^{22}\) or \(\mathcal{E} \in \mathbb{R}^{22 \times 22}\). We analyze \(\mathbb{D}(\mathcal{E})\) using Euclidean distance, while \(\mathcal{E}\) is compared using the metric defined in equation~\ref{eq:distance}. Consequently, commonly used time-frequency or time-domain features do not yield a directly comparable fixed-dimensional representation that captures articulatory structure in the same way.


\subsection{$\mathcal{H}$ can linearly map to $\mathbb{D}(\mathcal{E})$}
We test whether there exists a linear mapping defined by a weight matrix \( W \) and bias \( b \) such that \(\mathbb{D}(\mathcal{E}) \approx W\mathcal{H} + b\) with a high correlation%
\footnote{We actually aim to probe whether \(\mathcal{H}\) (\(768\)-\(1024\) dimensions) can map to \(\texttt{vec}(\mathcal{E})\) (\(961\) dimensions). However, the resulting \(\sim 10^6\)-parameter linear transformation would be severely ill-posed . To make this analysis tractable, we use \(\mathbb{D}(\mathcal{E})\) as a proxy because it provides a compact, well-conditioned, and physically meaningful representation grounded in articulatory mechanisms, making it well suited for linear probing. Importantly, this substitution is justified because both \(\mathcal{E}\) and \(\mathbb{D}(\mathcal{E})\) encode structured articulatory information, and the latter serves as a low-dimensional surrogate for the former, as shown in section~\ref{sec:tSNE}.}.

We use the training set described in section~\ref{sec:Data}  $\textsc{Data}_{\textsc{ general}}$ to learn this mapping and evaluate it on the corresponding test set. We report the Pearson correlation between the predicted sequences \(\mathbb{D}(\mathcal{E}')\) and the ground-truth \(\mathbb{D}(\mathcal{E})\) on the test set. The representations \(\mathcal{H}\) are extracted using \textsc{HuBERT} \citep{hsu2021hubert}, \textsc{Wav2Vec 2.0} \citep{baevski2020wav2vec}, and \textsc{WavLM} \citep{chen2022wavlm}. We evaluate \textsc{base} models with latent space dimension of 768 and 12 transformer layers, \textsc{large} models with latent space dimension of 1024 and 24 transformer layers, and \textsc{fine-tuned (ft)} models that have been tuned  for automatic speech recognition (ASR).

\begin{figure}[ht] 
    \centering 
    \includegraphics[width=0.48\textwidth]{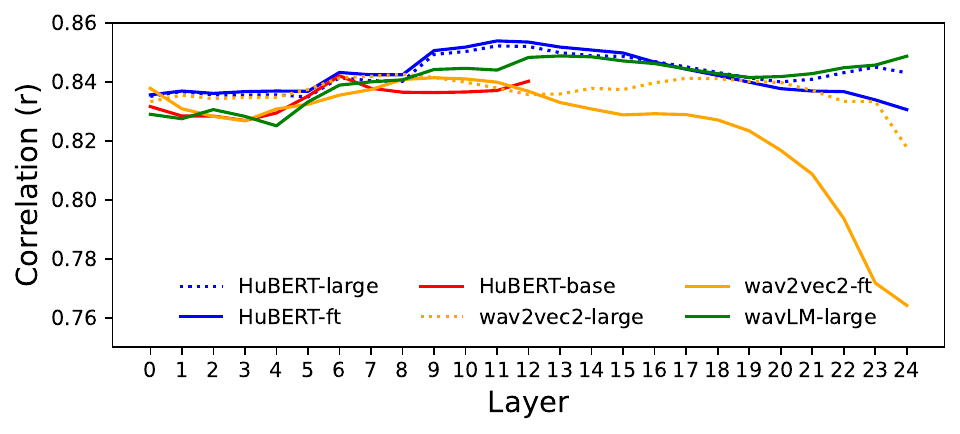} 
    \caption{Layer-wise correlation (\( r \)) between \(\mathbb{D}(\mathcal{E'})\) and \(\mathbb{D}(\mathcal{E})\) across different self-supervised speech models. A simple linear mapping is used to predict \(\mathbb{D}(\mathcal{E'})\) from \(\mathcal{H}\).}

    \label{fig:corrEMGcovAudioSSL} 
\end{figure}

Correlation coefficients (\emph{r}) across models and layers are shown in figure~\ref{fig:corrEMGcovAudioSSL}. We find that a simple linear model can predict \(\mathbb{D}(\mathcal{E})\) from \(\mathcal{H}\) with a correlation as high as \( r = 0.85 \). The layer-wise trends across different models partially mirror the observations reported in \citep{cho2023evidence, cho2024self} for electromagnetic articulography (EMA), where two local peaks were consistently observed across models. In our case, we observe two local peaks for \textsc{Wav2Vec 2.0} models but only a single dominant peak for \textsc{HuBERT} and \textsc{WavLM} models. A sharp decline in correlation emerges in the upper layers of fine-tuned models, reflecting the growing influence of task-specific objectives. This effect is especially pronounced for \textsc{Wav2Vec 2.0} compared to \textsc{HuBERT} and \textsc{WavLM}. 

Notably, for the \textsc{HuBERT-base} model, the peak correlation at layer~6 aligns with the layer previously identified as optimal for discrete speech resynthesis and spoken language modeling~\citep{lakhotia-etal-2021-generative}. While prior work established this empirical result, the mechanistic basis for this peak remained unclear. Our analysis provides a principled interpretation: layer~6 exhibits the strongest linear predictive power for \(\mathbb{D}(\mathcal{E})\), which encodes structured and discriminative articulatory information (i.e., different articulatory gestures such as tongue and jaw positions naturally form separable clusters). This tight alignment between articulatory structure and model representations offers a direct explanation for why layer~6 is particularly effective for downstream speech resynthesis and language modeling. In short, the layer that best captures articulatory mechanisms is also the one that yields the strongest downstream performance, providing convergent evidence for its functional role.

\begin{figure}[ht] 
    \centering 
    \includegraphics[width=0.48\textwidth]{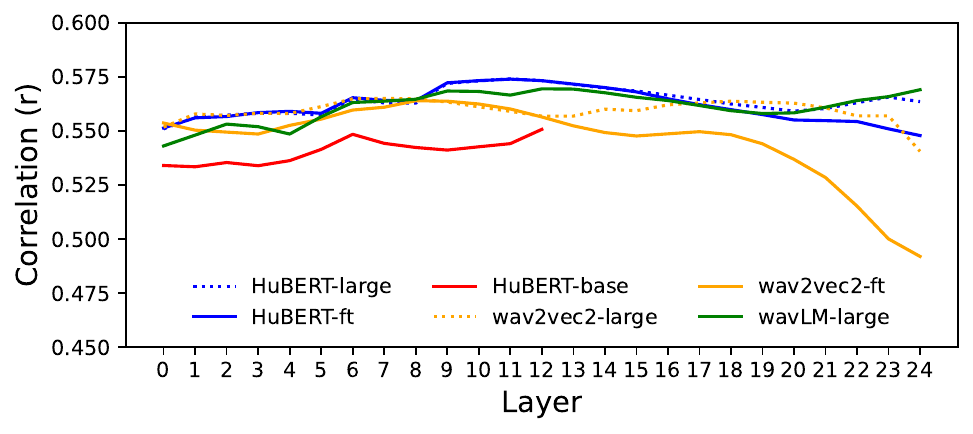} 
    \caption{Layer-wise correlation (\( r \)) between $\mathcal{B'}$ and \(\mathcal{B}\) across different self-supervised speech models. A simple linear mapping is used to predict $\mathcal{B'}$ from \(\mathcal{H}\).}

    \label{fig:corrEMGspectraAudioSSL} 
\end{figure}

We also examine whether a similar linear mapping exists between EMG spectrogram features (\(\texttt{vec}(\mathcal{B})\)) and \(\mathcal{H}\). Frequency bands of \(\mathcal{B}\) are obtained using five frequency bins, as described in section~\ref{sec:Methods}. However, the resulting correlation coefficients are substantially lower, with a maximum correlation of approximately \( r = 0.57 \) (figure~\ref{fig:corrEMGspectraAudioSSL}). For comparison, we also compute correlations for linear mapping between \(\mathcal{A}\) and \(\mathbb{D}(\mathcal{E})\) (\( r = 0.61 \)), which is considerably lower than the correlation between \(\mathcal{H}\) and \(\mathbb{D}(\mathcal{E})\).

The above observations indicate that among the different EMG feature representations considered, \(\mathbb{D}(\mathcal{E})\) exhibits the strongest linear alignment with the self-supervised speech feature space \(\mathcal{H}\). This strong correspondence suggests that \(\mathbb{D}(\mathcal{E})\) (consequently, \(\texttt{vec}(\mathcal{E})\)) and \(\mathcal{H}\) encode highly compatible representations, making them particularly well suited for EMG-to-audio learning. In contrast, EMG spectrogram features ($\mathcal{B}$) and their alignment with (\(\mathcal{H}\)) yield notably weaker correlations.

\subsection{emg2speech synthesis}

As shown earlier, the following relationship holds:
\[
\begin{aligned}
\mathcal{H} &\xrightarrow{\texttt{linear mapping}} \mathbb{D}(\mathcal{E})
\xrightarrow{\texttt{gesture-specific clustering}} \\
&\textsc{orofacial movements}.
\end{aligned}
\]

The existence of a simple linear mapping from \(\mathcal{H}\) to \(\mathbb{D}(\mathcal{E})\) is informative because it indicates that the self-supervised representations \(\mathcal{H}\) encode articulatory structure consistent with underlying muscle activations. This forward mapping is well posed: \(\mathcal{H}\) has moderate dimensionality (\(768\)–\(1024\)), whereas \(\mathbb{D}(\mathcal{E})\) is low dimensional (\(31\)), allowing the mapping to be estimated stably using linear regression.

By contrast, the inverse direction \(\mathbb{D}(\mathcal{E}) \rightarrow \mathcal{H}\) is intrinsically underdetermined and not uniquely invertible in the linear setting, since multiple high-dimensional speech representations can correspond to the same low-dimensional articulatory configuration. This challenge is further compounded when temporal alignment between EMG and audio is unknown. Nevertheless, the existence of a reliable forward mapping suggests that recovering \(\mathcal{H}\) from EMG is a feasible learning problem when using an appropriate nonlinear mapping
\footnote{For linear probing, we use low-dimensional versions of both covariance and spectrogram representations: \(\mathbb{D}(\mathcal{E})\) and a 5-bin spectrogram \(\mathcal{B}\). For speech synthesis, we instead use full-resolution features (\(\texttt{vec}(\mathcal{E})\) and a 31-bin \(\mathcal{B}\)) to preserve fine-grained cross-channel and spectral structure.}.

\begin{figure}[ht] 
    \centering 
    \includegraphics[width=0.47\textwidth]{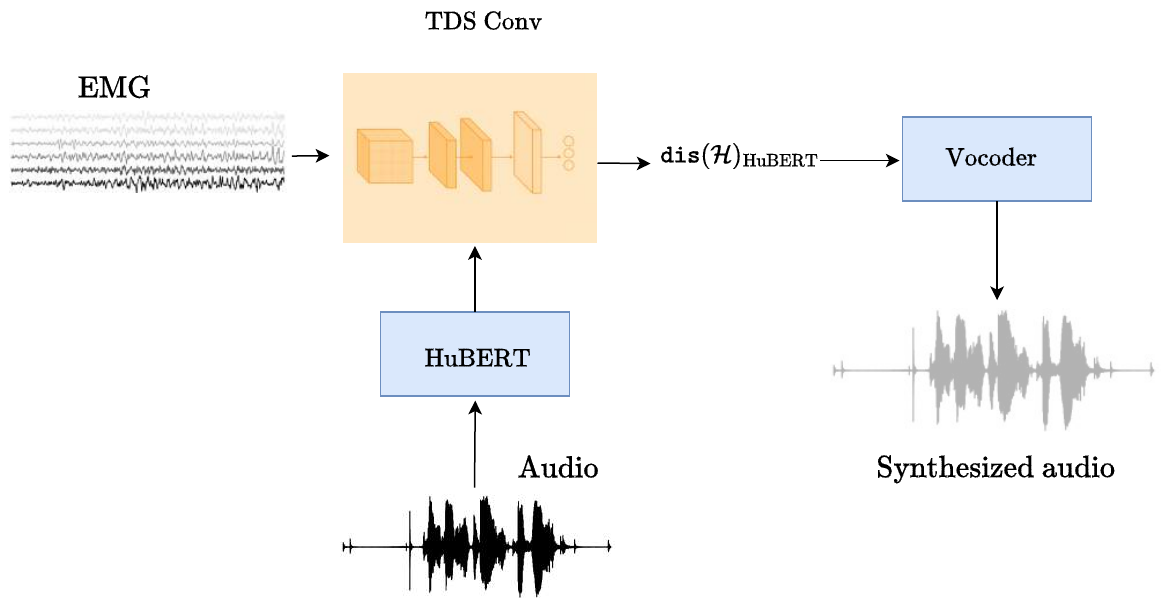} 
    \caption{Multivariate EMG signals are converted into \(\texttt{vec}(\mathcal{E})\), \(\mathbb{D}(\mathcal{E})\), or \(\mathcal{B}\), and then passed through a \textsc{TDS Conv} block to predict \(\texttt{dis}(\mathcal{H})_{\textsc{HuBERT}}\), which are subsequently fed into a vocoder to synthesize audio. Frozen neural network components are shown in \textcolor{blue}{blue}, and trainable components are shown in \textcolor{orange}{orange}.}
    \label{fig:emgSpeech} 
\end{figure}

Motivated by this observation, we consider alignment-free prediction of \(\mathcal{H}\)-derived representations from EMG features (\(\texttt{vec}(\mathcal{E})\), \(\mathbb{D}(\mathcal{E})\), or \(\texttt{vec}(\mathcal{B})\)). Because the inverse linear mapping is ill posed, we model it using a nonlinear sequence-to-sequence architecture that can capture temporal and contextual dependencies present in \(\mathcal{H}\). Concretely, EMG features are provided as input to a TDS convolutional network (section~\ref{sec:seq2seq}) that predicts discrete units associated with \(\mathcal{H}\). We use the 100-unit discretization from layer 6 of \textsc{HuBERT-base}~\citep{lakhotia-etal-2021-generative}, denoted \(\texttt{dis}(\mathcal{H})_{\textsc{HuBERT}}\). Training is performed with the connectionist temporal classification (CTC) loss~\citep{graves2006connectionist}, which enables learning without explicit frame-level alignment between EMG sequences and \(\texttt{dis}(\mathcal{H})_{\textsc{HuBERT}}\). Finally, the predicted \(\texttt{dis}(\mathcal{H})_{\textsc{HuBERT}}\) sequence is converted to audio using a pretrained Tacotron vocoder~\citep{Tacotron}. The overall architecture is illustrated in figure~\ref{fig:emgSpeech}.

\subsubsection{Results for $\textsc{Data}_{\textsc{ general}}$}
\label{sec:results_gen}

We use Google text-to-speech (gTTS) to synthesize audio from the corresponding text transcripts. From this synthesized audio, we extract the discrete HuBERT units \(\texttt{dis}(\mathcal{H})_{\textsc{HuBERT}}\)\footnote{For both \(\textsc{Data}_{\textsc{ general}}\) and \(\textsc{Data}_{\textsc{ ALS}}\), we do not use subject-recorded audio for EMG-to-speech synthesis. The healthy participant vocalized the sentences during recording, whereas the ALS participant articulated them silently. In both cases, we rely only on transcript-based gTTS audio to derive \(\texttt{dis}(\mathcal{H})_{\textsc{HuBERT}}\). Subject-recorded audio from \(\textsc{Data}_{\textsc{ general}}\) is used only for probing linearity in the previous section. This design is motivated by clinical scenarios in which parallel EMG and audio recordings may be unavailable. Note that the gTTS audio is not temporally aligned with the EMG, which makes the translation problem more challenging than settings with paired, time-synchronized supervision.}. We report \(\texttt{dis}(\mathcal{H})_{\textsc{HuBERT}}\) decoding results in table~\ref{tab:hubResults}.

We provide \(\texttt{vec}(\mathcal{E})\), \(\mathbb{D}(\mathcal{E})\), or \(\texttt{vec}(\mathcal{B})\) as input to the TDS network, and train it to predict the corresponding \(\texttt{dis}(\mathcal{H})_{\textsc{HuBERT}}\) unit sequence using the CTC loss. For example, for the sentence \(_{\textsc{t-start}}<\textcolor{blue}{\textsc{It Was Paid For}}>_{\textsc{t-end}}\) with target \textsc{HuBERT} units \textcolor{green!30!black}{\textsc{\small 71-12-71-12-4-12-4-40-93-86-13-58-32-1-99-...}}, the model learns a mapping from the EMG feature sequence to the target unit sequence. During inference, the model outputs a probability distribution over all 100 \(\texttt{dis}(\mathcal{H})_{\textsc{HuBERT}}\) units at each time step, and we decode the most likely unit sequence using {\em greedy search}. For instance, the decoded sequence may be \textcolor{purple}{\textsc{71-12-57-4-54-40-93-86-13-58-16-14-76-6-36-...}}. We compute the unit error rate (UER) as the Levenshtein distance between the target and predicted \(\texttt{dis}(\mathcal{H})_{\textsc{HuBERT}}\) sequences, normalized by the target sequence length (see appendix~\ref{apd:ablation} for ablation studies on training data size
).

\begin{table}[h!]
\centering
\caption{Unit error rate (UER) for different EMG feature representations when predicting \(\texttt{dis}(\mathcal{H})_{\textsc{HuBERT}}\) units on $\textsc{Data}_{\textsc{ general}}$. The dataset and preprocessing details are described in section~\ref{sec:Data}. Lower UER is better. Values are averaged over 5 random seeds.}
\begin{tabular}{lcc}
\hline
\textsc{Model input} & \textsc{UER} (\footnotesize$\%\downarrow$) &
\shortstack{%
\rule{0pt}{0.5em}
\textsc{\scriptsize input feature}\\[0.25em]
\textsc{\scriptsize dimension}%
}\\
\hline

 $\texttt{vec}(\mathcal{B})$ & 62.71 $\pm$ 0.50\hspace{0.1cm} & 961 \\\hline
 $\mathbb{D}(\mathcal{E})$ & 62.16 $\pm$ 0.50\hspace{0.1cm} & 31\\\hline
 $\texttt{vec}(\mathcal{E})$ & \textbf{56.08 $\pm$ 0.91}\hspace{0.1cm} & 961\\\hline
\end{tabular}
\label{tab:hubResults}
\end{table}

We also present the results of phoneme-level decoding in table~\ref{tab:phoneResults}. For the sentence $_{\textsc{t-start}}<$\textcolor{blue}{\textsc{It Was Paid For}}$>_{\textsc{t-end}}$ with the corresponding phonemic transcription \textcolor{green!30!black}{\textsc{ih-t $\textsc{\tiny space}$ w-aa-z $\textsc{\tiny space}$ p-ey-d $\textsc{\tiny space}$ f-ao-r}}, the TDS model is trained to learn the mapping from \(\texttt{vec}(\mathcal{E})\), \(\mathbb{D}(\mathcal{E})\), or \(\texttt{vec}(\mathcal{B})\) to phoneme sequences using the CTC loss. During inference, the model outputs probabilities for all 40 English phonemes at each time step, and the predictions are decoded using {\em greedy search}. For example, the decoded output might be \textcolor{purple}{\textsc{ih-t $\textsc{\tiny space}$ w-aa-z $\textsc{\tiny space}$ p-ey-t $\textsc{\tiny space}$ f-ao-r}}. We compute the phoneme error rate (PER) as the Levenshtein distance between the target and decoded phoneme sequences, normalized by the length of the target sequence.

\begin{table}[h!]
\centering
\caption{Phoneme error rate (PER) for different EMG feature representations when predicting phonemes on $\textsc{Data}_{\textsc{ general}}$. The dataset and preprocessing details are described in section~\ref{sec:Data}. Lower PER is better. Values are averaged over 5 random seeds.}
\begin{tabular}{lcc}
\hline
\textsc{Model input} & \textsc{PER} (\footnotesize$\%\downarrow$) &
\shortstack{%
\rule{0pt}{0.5em}
\textsc{\scriptsize input feature}\\[0.25em]
\textsc{\scriptsize dimension}%
}\\
\hline
 $\texttt{vec}(\mathcal{B})$ & 44.40 $\pm$ 2.28\hspace{0.1cm} & 961\\\hline
 $\mathbb{D}(\mathcal{E})$ & 44.40 $\pm$ 1.51\hspace{0.1cm} & 31\\\hline
 $\texttt{vec}(\mathcal{E})$ & \textbf{32.78 $\pm$ 0.66} & 961\hspace{0.1cm} \\\hline
\end{tabular}

\label{tab:phoneResults}
\end{table}

As shown in tables~\ref{tab:hubResults} and \ref{tab:phoneResults}, \(\texttt{vec}(\mathcal{E})\) outperforms \(\texttt{vec}(\mathcal{B})\). \(\mathcal{B}\) was computed using 31 linearly spaced frequency bins, and for any given time frame, both \(\texttt{vec}(\mathcal{E})\) and \(\texttt{vec}(\mathcal{B})\) have 961 dimensions. Notably, even \(\mathbb{D}(\mathcal{E})\), which has only 31 dimensions (i.e., 31$\times$ fewer dimensions than  \(\texttt{vec}(\mathcal{B})\)), performs on par with \(\texttt{vec}(\mathcal{B})\). This finding is consistent with the linear mapping results shown in figures~\ref{fig:corrEMGcovAudioSSL} and \ref{fig:corrEMGspectraAudioSSL}.

\textbf{\textsc{Phoneme guided decoding of \(\texttt{dis}(\mathcal{H})_{\textsc{HuBERT}}\)}}: as shown in table~\ref{tab:phoneResults}, phoneme sequences can be decoded more accurately than
\(\texttt{dis}(\mathcal{H})_{\textsc{HuBERT}}\) unit sequences. Motivated by this observation, we train the TDS
convolutional encoder in figure~\ref{fig:emgSpeech} with two prediction heads: one for phonemes,
producing framewise posteriors \(P(\textsc{phoneme}\mid \textsc{EMG})\), and one for
\(\texttt{dis}(\mathcal{H})_{\textsc{HuBERT}}\) units, producing framewise posteriors
\(P(\texttt{dis}(\mathcal{H})_{\textsc{HuBERT}}\mid \textsc{EMG})\). The model is optimized with CTC losses for
both outputs, together with an additional consistency loss that encourages phoneme-consistent
\(\texttt{dis}(\mathcal{H})_{\textsc{HuBERT}}\) predictions. Specifically, we use a precomputed lookup table
\(P(\textsc{phoneme}\mid \text{\em u})\), where \(\text{\em u} \in \mathcal{U}\) and \(\mathcal{U}=\texttt{dis}(\mathcal{H})_{\textsc{HuBERT}}\)
denotes the discrete HuBERT unit set, to transform the unit posterior at each frame \(t\) into a phoneme
distribution by marginalizing over units:
\[
{\footnotesize
\begin{aligned}
\tilde P(\textsc{phoneme}\mid \textsc{EMG})_t
&= \sum_{\text{\em u} \in \mathcal{U}} P(\textsc{phoneme}\mid \text{\em u})\,P(\text{\em u}\mid \textsc{EMG})_t .
\end{aligned}
}
\]
We then minimize a cross-entropy loss between \(\tilde P(\textsc{phoneme}\mid \textsc{EMG})_t\) and the
phoneme-head posterior \(P(\textsc{phoneme}\mid \textsc{EMG})_t\) (see appendix~\ref{apd:phoneGuided} for more
details). At inference time, we decode only \(\texttt{dis}(\mathcal{H})_{\textsc{HuBERT}}\) units from the unit head using greedy decoding. Resulting improvement is shown in table~\ref{tab:hubGuidedAblation}. 
\begin{table}[h!]
\centering
\caption{Effect of phoneme-guided training on \textsc{HuBERT} unit decoding with \(\texttt{vec}(\mathcal{E})\) as input on $\textsc{Data}_{\textsc{ general}}$. Values are averaged over 5 random seeds. The improvement is statistically significant ({\em p }$< \text{ 10}^{-6}$).}
\begin{tabular}{lc}
\hline
\textsc{Training objective} & \textsc{UER} (\footnotesize$\%\downarrow$)\\
\hline
Unit CTC only                         & 56.08 $\pm$ 0.91 \\\hline
Phoneme-guided decoding & \textbf{51.81 $\pm$ 0.62} \\
\hline
\end{tabular}
\label{tab:hubGuidedAblation}
\end{table}

Furthermore, three human raters (see appendix \ref{apd:human}) listened to all 400 synthesized audio samples in the test set (generated from \(\texttt{dis}(\mathcal{H})_{\textsc{HuBERT}}\) units obtained via phoneme-guided decoding with \(\texttt{vec}(\mathcal{E})\) as input) and transcribed them. We compute the word error rate (WER) as the Levenshtein distance between each rater's transcription and the ground-truth transcript, normalized by the length of the ground-truth transcript. We report the resulting WERs in table~\ref{tab:WER_both}.
\subsubsection{Results for $\textsc{Data}_{\textsc{ ALS}}$}
We follow the same preprocessing, model architecture, and training procedure used for $\textsc{Data}_{\textsc{ general}}$. We report unit decoding performance in table~\ref{tab:hubResultsALS}. To assess end-to-end intelligibility, we synthesize speech from the decoded units and measure word error rate (WER) using human transcriptions (on all 60 synthesized audios in the test set); the resulting WER is reported in table~\ref{tab:WER_both}.

\begin{table}[h!]
\centering
\caption{Unit error rate (UER) for different EMG feature representations when predicting \(\texttt{dis}(\mathcal{H})_{\textsc{HuBERT}}\) units on $\textsc{Data}_{\textsc{ ALS}}$. Dataset and preprocessing details are described in section~\ref{sec:Data}. Lower UER is better. Values are averaged over 5 random seeds.}
\begin{tabular}{lc}
\hline
\textsc{Model input} & \textsc{UER} (\footnotesize$\%\downarrow$)\\
\hline
$\texttt{vec}(\mathcal{B})$ & 59.18 $\pm$ 2.81\vspace{0.1cm} \\\hline
$\mathbb{D}(\mathcal{E})$ & 54.27 $\pm$ 0.56\vspace{0.1cm} \\\hline
$\texttt{vec}(\mathcal{E})$ & 52.29 $\pm$ 0.91\vspace{0.1cm} \\\hline
$\texttt{vec}(\mathcal{E})$ with \\ phoneme-guided decoding & \textbf{46.98 $\pm$ 1.11}\vspace{0.1cm} \\\hline
\end{tabular}
\label{tab:hubResultsALS}
\end{table}

\begin{table}[h!]
\centering
\caption{WER as rated by human transcribers.}
\setlength{\tabcolsep}{4pt} 
\renewcommand{\arraystretch}{1.05} 
\begin{tabular}{lcc}
\hline
\textsc{\footnotesize Human} & \textsc{\footnotesize Data$_\textsc{ General}$} & \textsc{\footnotesize Data$_\textsc{ ALS}$} \\
\textsc{\footnotesize transcriber} & \textsc{\footnotesize WER} (\footnotesize$\%\downarrow$) & \textsc{\footnotesize WER} (\footnotesize$\%\downarrow$) \\
\hline
1 & 65.09 & 52.69 \\
2 & 59.32 & 50.97 \\
3 & 63.23 & 48.81 \\
\hline
Average & \textbf{62.55 $\pm$ 2.40} & \textbf{50.82 $\pm$ 1.59} \\
\hline
\end{tabular}
\label{tab:WER_both}
\end{table}

We contextualize these WERs relative to prior brain-computer interface studies in appendix~\ref{apd:Lit}. 
Please see the following links for demonstrations:
\textcolor{magenta}{\Large \faVolumeUp}\,\href{https://harshavardhanatg.github.io/emg2speech.github.io/}{Audio}.
\quad We also provide the ground-truth transcripts and test-set transcripts from three human raters:
\textcolor{magenta}{\Large \faBook}\,\href{https://harshavardhanatg.github.io/emg2speech.github.io/static/pdfs/allTranscripts.html}{Transcriptions}.

\section{Conclusion}
We present methods and datasets that convert orofacial EMG signals directly into speech, and we demonstrate the system with both a healthy participant and a participant diagnosed with ALS.

\newpage

\section{Limitations}
This work primarily focuses on the technical aspects of EMG-to-speech modeling, including characterizing the structure of orofacial EMG signals, quantifying their relationship to self-supervised speech representations, and designing encoder architectures grounded in articulatory mechanisms. Our clinical demonstration uses approximately one hour of data from a single participant with ALS. As a result, we do not yet characterize how performance evolves under day-to-day distribution shifts in EMG signals (e.g., changes in electrode placement, skin impedance, fatigue, or disease progression). Consequently, we also do not evaluate whether this non-stationarity is more or less challenging than the distribution shifts observed in invasive neural interfaces \citep{wairagkar2025instantaneous, willett2023high}.

In addition, we do not demonstrate sustained, long-term performance of this non-invasive neuroprosthesis. In contrast, prior work on invasive neuroprostheses has reported stability over extended periods in related decoding settings, including brain-to-text \citep{fan2023plug} and cursor-based brain-computer interfaces \citep{wilson2025long}. 

Finally, we do not explore whether large-scale pretrained EMG-to-speech models can improve decoding performance. For example, in related work on EMG-based keyboard typing (\textsc{emg2qwerty}) \citep{sivakumaremg2qwerty}, pretraining on data from 100 individuals improved accuracy after fine-tuning to new individuals, although zero-shot performance remained limited. Speech may be more challenging than discrete key typing, and future work should investigate how to build and effectively leverage large-scale pretrained models for EMG-to-speech translation. 

We are actively addressing these limitations through ongoing longitudinal studies and by expanding data collection to build larger EMG-to-speech corpora from individuals with diverse clinical etiologies, including ALS and laryngectomy.

\section{Ethical considerations}
\label{sec:ethics}
Research was conducted in accordance with the principles embodied in the Declaration of Helsinki and in accordance with the University of California, Davis Institutional Review Board (IRB) Administration protocol 2078695-1. All participants provided written informed consent. All participants also provided consent for publication of deidentified data. Volunteers of any gender and from all racial and ethnic groups were eligible to participate. Participants were required to be at least 18 years old, able to understand spoken and written English, and able to follow task instructions. Participants had no skin conditions or wounds at electrode placement sites and were excluded if they had uncorrected vision problems. Children, individuals unable to provide informed consent, and prisoners were not included in the experiments. All participants were compensated in accordance with IRB protocols.

The participant with ALS was first diagnosed in 2019 and has non-familial ALS with spasticity. 

\section*{Acknowledgments}

This work was supported by awards to Lee M. Miller from: Accenture, through the Accenture Labs Digital Experiences group; CITRIS and the Banatao Institute at the University of California; the University of California Davis School of Medicine (Cultivating Team Science Award); the University of California Davis Academic Senate; a UC Davis Science Translation and Innovative Research (STAIR) Grant; and the Child Family Fund for the Center for Mind and Brain. 

Harshavardhana T. Gowda is supported by Neuralstorm Fellowship, NSF NRT Award No. 2152260 and Ellis Fund
administered by the University of California, Davis.

We thank Karen Dhillon and Craig M. McDonald's Neuromuscular Research Lab at UC Davis for valuable guidance and participant recruitment.

\section*{Conflict of interest}

H. T. Gowda and L. M. Miller are inventors on intellectual property related to {\em silent} speech owned by the Regents of University of California, not presently licensed.

\section*{Author contributions}
\begin{itemize}
    \item Harshavardhana T. Gowda: Conceptualization, mathematical formulation, method development, data analysis, experimental design, data collection software development, data collection, and manuscript preparation.
    \item Daniel C. Comstock: Data collection and manuscript review.
    \item Lee M. Miller: Conceptualization, funding, participant recruitment, and manuscript review.
\end{itemize}

\appendix

\section{Experimental details}
\label{appendix:exp}

\begin{figure*}[t]
  \centering

  \begin{subfigure}[t]{0.32\textwidth}
    \includegraphics[width=\linewidth, trim=0 25 0 38, clip]{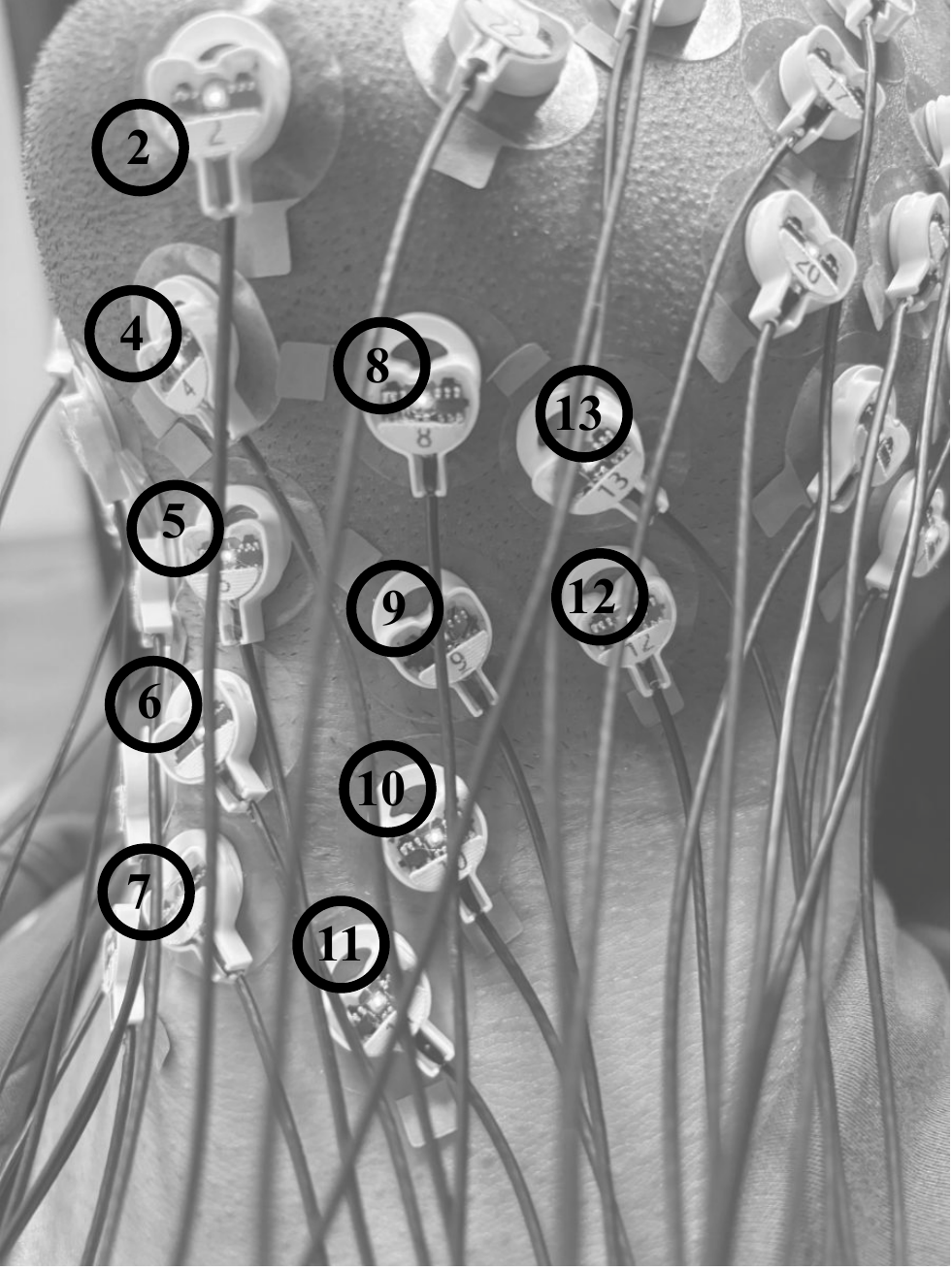} 
  \end{subfigure}
  \hfill
  \begin{subfigure}[t]{0.32\textwidth}
    \includegraphics[width=\linewidth, trim=0 0 0 0, clip]{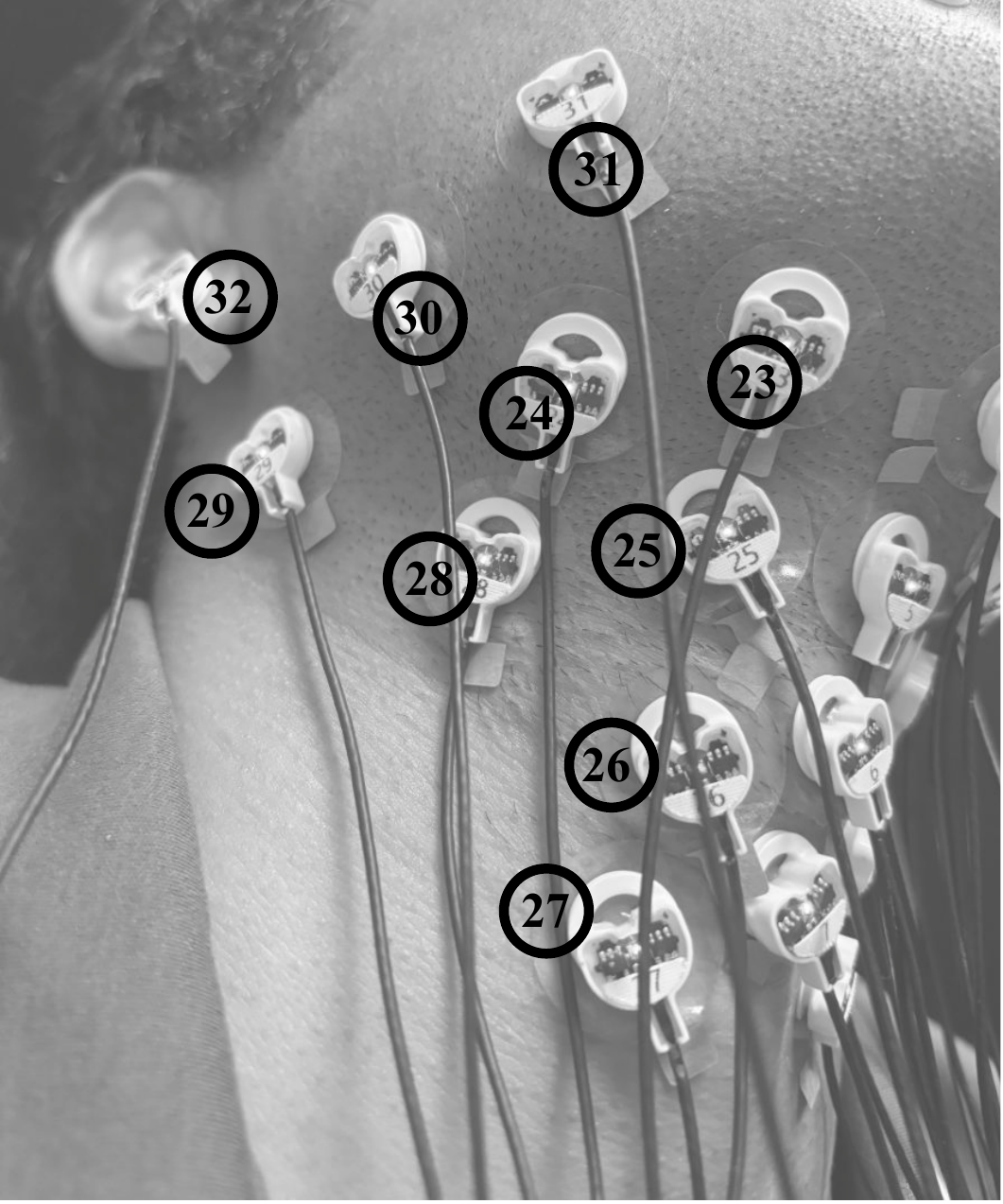} 
  \end{subfigure}
  \hfill
  \begin{subfigure}[t]{0.32\textwidth}
    \includegraphics[width=\linewidth, trim=0 20 0 75, clip]{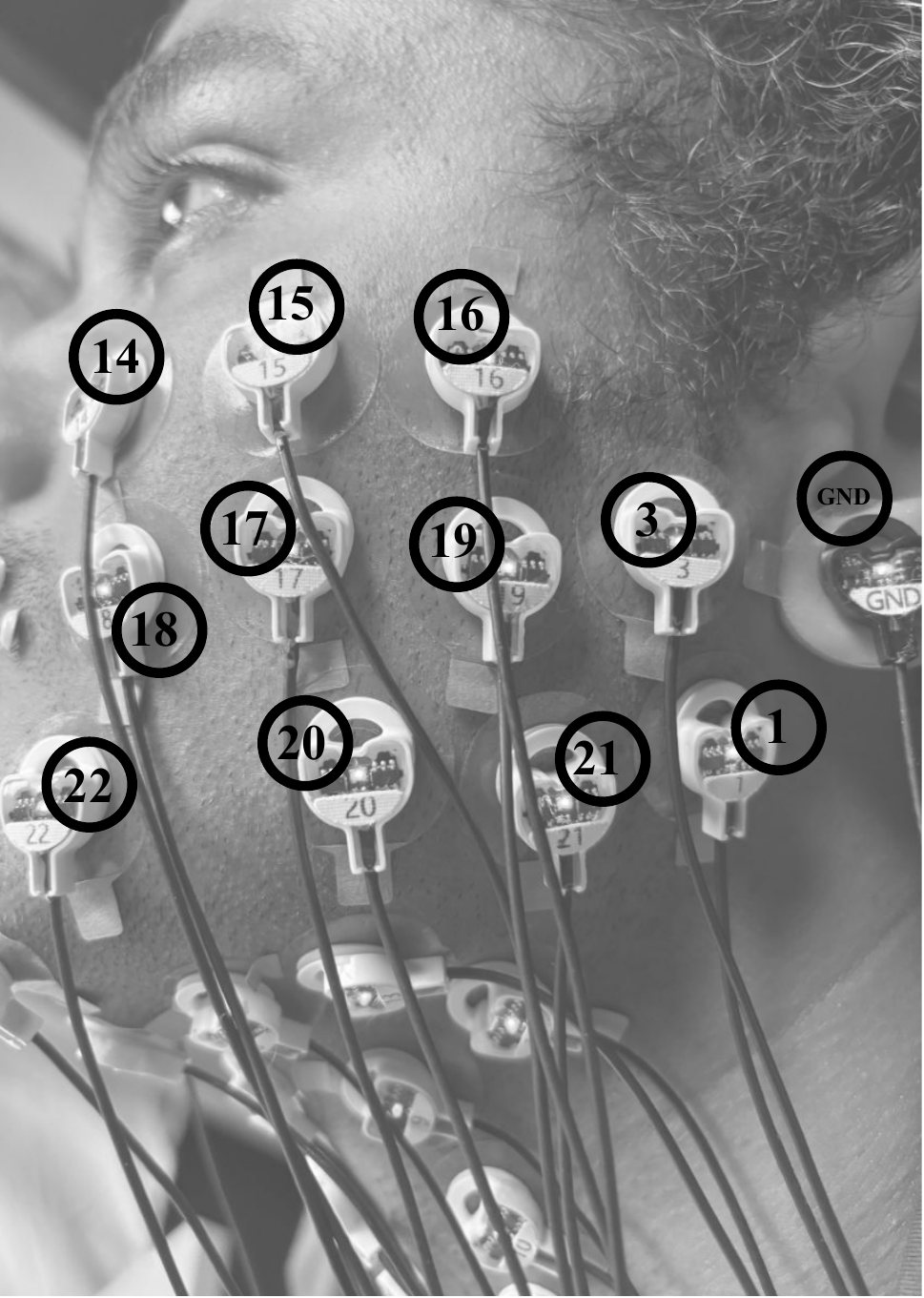} 
  \end{subfigure}

  \caption{\textsc{Left:} electrode placement on the left side of the neck. 
  \textsc{Middle:} electrode placement on the right side of the neck. 
  \textsc{Right:} electrode placement on the left cheek.}
  \label{fig:electrodePlacement}
\end{figure*}

We collect EMG signals from 31 sites on the neck, chin, jaw, cheek, and lips using monopolar electrodes.
An \textsc{actiCHamp Plus} amplifier and associated active electrodes from \textsc{BrainVision}
(\href{https://brainvision.com/products/actichamp-plus/}{Brain Vision}) are used to record EMG signals at 5000~Hz.
To ensure proper contact between the skin surface and electrodes, we use \textsc{SuperVisc}, a high-viscosity
electrolyte gel from \textsc{Easycap} (\href{https://shop.easycap.de/products/supervisc}{Easycap}).
We develop a software suite in a \textsc{Python} environment to provide visual cues to participants and to collate
and store timestamped data. For time synchronization, we use Lab Streaming Layer
(\href{https://labstreaminglayer.org}{LSL}). See figure~\ref{fig:electrodePlacement} for electrode placement.
In addition to the 31 data electrodes, we also use a \textsc{ground} electrode (marked as \textsc{gnd}) and a
\textsc{reference} electrode (marked as 32). The \textsc{ground} electrode is placed on the left earlobe and the
\textsc{reference} electrode is placed on the right earlobe.

Before signal acquisition, participants are briefed on the experimental protocol and seated comfortably in a chair.
Sentence start and end times are timestamped using mouse clicks. When a participant is ready to articulate a
sentence, they click the mouse to prompt the sentence to appear on the screen. Once articulation is complete, they
click again to indicate the end, which causes the sentence to disappear. This allows participants to articulate at
their own pace.

The data collection environment is carefully controlled to reduce AC electrical interference. EMG signals undergo
minimal preprocessing. The signal from the \textsc{reference} channel (electrode 32) is subtracted from all other
channels. The resulting signals are bandpass filtered using a third-order Butterworth filter between 80 and
1000~Hz and segmented according to sentence start and end times based on synchronized timestamps. The segmented
sentences are subsequently $z$-normalized along the time dimension for each channel.

The electrodes are positioned over regions that overlay muscle groups involved in speech articulation, providing
coverage of key articulators such as the tongue, jaw, lips, and larynx. Electrode locations 19, 21, 3, and 1
approximately overlie the \text{\em hyoglossus}, \text{\em palatoglossus}, and \text{\em styloglossus} muscles.
These muscles are located in the lower cheek region and play a vital role in tongue movement. They are also
consistently recruited across a wide range of articulatory gestures. Muscles in the upper and posterior cheek
regions include the \text{\em masseter} and \text{\em temporalis}, which control jaw motion, and the
\text{\em zygomaticus}, which is involved in upper lip elevation. These muscles correspond approximately to electrode
regions around nodes 22, 18, 17, and 15 in figure~\ref{fig:electrodePlacement}. Electrodes located beneath the jaw
capture activity from muscles involved in tongue protrusion and jaw-tongue coordination, such as the
\text{\em genioglossus} near electrodes 8, 9, 23, and 25, as well as the \text{\em digastric}. Additionally, electrodes
near the laryngeal region, including nodes 6, 7, 10, 11, 26, and 27, reflect activity from muscles that modulate
laryngeal and hyoid position, such as the \text{\em sternohyoid}, \text{\em stylohyoid}, and \text{\em digastric}.
These muscles contribute to pitch control, vowel shaping, and jaw movement.

\section{Detailed explanation: phoneme guided decoding of \(\texttt{dis}(\mathcal{H})_{\textsc{HuBERT}}\)}
\label{apd:phoneGuided}
We train a bidirectional gated recurrent unit (GRU) model to map
\(\texttt{dis}(\mathcal{H})_{\textsc{HuBERT}}\) unit sequences to phoneme sequences using a CTC objective
on \(\textsc{Data}_{\textsc{ general}}\) (train-val-test split as described in section~\ref{sec:Data}).
After CTC decoding, the model achieves \(\mathrm{PER}=0\%\) (phoneme error rate) on the corresponding test split.
Using this trained model, we estimate a unit-to-phoneme conditional table by aggregating framewise posteriors:
for each HuBERT unit \(\text{\em u}\), we collect the predicted phoneme distribution
\(P_{\textsc{gru}}(\textsc{phoneme}_{t}\mid \text{\em u})\) at every frame \(t\) where \(\text{\em u}_t=\text{\em u}\),
and average across all such frames, i.e.,
\[
\begin{aligned}
& P(\textsc{phoneme} \mid \text{\em u}) = \\
& \qquad \frac{1}{N_\text{\em u}}\sum_{t:\, \text{\em u}_t=\text{\em u}}
P_{\textsc{gru}}\!\left(\textsc{phoneme}_t = \textsc{phoneme} \mid \text{\em u}\right),
\end{aligned}
\]
where \(N_\text{\em u} = \lvert\{t : \text{\em u}_t = \text{\em u}\}\rvert\) denotes the total number of frames
in the dataset for which the \textsc{HuBERT} unit equals \(\text{\em u}\).
We remove the CTC blank symbol from \(P(\cdot \mid \text{\em u})\) and renormalize.
This yields \(P(\textsc{phoneme}\mid \texttt{dis}(\mathcal{H})_{\textsc{HuBERT}})\), which we use as a fixed
probabilistic mapping in our consistency regularization.\\

For phoneme-guided decoding of \(\texttt{dis}(\mathcal{H})_{\textsc{HuBERT}}\), we train the TDS convolutional
encoder in figure~\ref{fig:emgSpeech} with two heads: one for phonemes, producing framewise posteriors
\(P(\textsc{phoneme}\mid \textsc{EMG})\), and one for \(\texttt{dis}(\mathcal{H})_{\textsc{HuBERT}}\) units,
producing framewise posteriors \(P(\text{\em u}\mid \textsc{EMG})\), where \(\text{\em u}\in\mathcal{U}\) and
\(\mathcal{U}=\texttt{dis}(\mathcal{H})_{\textsc{HuBERT}}\) denotes the discrete HuBERT unit set.
The model is optimized with CTC losses for both outputs. Additionally, we impose a consistency loss by using the
precomputed lookup table \(P(\textsc{phoneme}\mid \text{\em u})\) to transform the unit posterior at each frame \(t\)
into a phoneme distribution via marginalization over units:
\[
\begin{aligned}
\tilde P(\textsc{phoneme}\mid \textsc{EMG})_t
&= \sum_{\text{\em u}\in\mathcal{U}} P(\textsc{phoneme}\mid \text{\em u}) \\
&\qquad \cdot P(\text{\em u}\mid \textsc{EMG})_t .
\end{aligned}
\]

We then minimize a cross-entropy loss between \(\tilde P(\textsc{phoneme}\mid \textsc{EMG})_t\) and the phoneme-head
posterior \(P(\textsc{phoneme}\mid \textsc{EMG})_t\):
\[
\begin{aligned}
\mathcal{L}_{\text{cons}}
&= -\sum_{t}\sum_{\textsc{phoneme}}
P(\textsc{phoneme}\mid \textsc{EMG})_t \\
&\qquad \cdot \log \tilde P(\textsc{phoneme}\mid \textsc{EMG})_t.
\end{aligned}
\]
The total training objective is a weighted sum of the two CTC losses and the proposed consistency term:
\[
\mathcal{L}_{\text{total}}
\;=\;
\lambda_{\text{unit}}\,\mathcal{L}_{\text{CTC}}^{\text{unit}}
\;+\;
\lambda_{\text{phone}}\,\mathcal{L}_{\text{CTC}}^{\text{phone}}
\;+\;
\lambda_{\text{cons}}\,\mathcal{L}_{\text{cons}}.
\]
In our experiments, we use $\lambda_{\text{unit}}$ = 0.8, $\lambda_{\text{phone}}$ = 0.1, and $\lambda_{\text{cons}}$ = 0.1.

\begin{figure*}[!t]  
  \centering
  \includegraphics[width=1.2\textwidth]{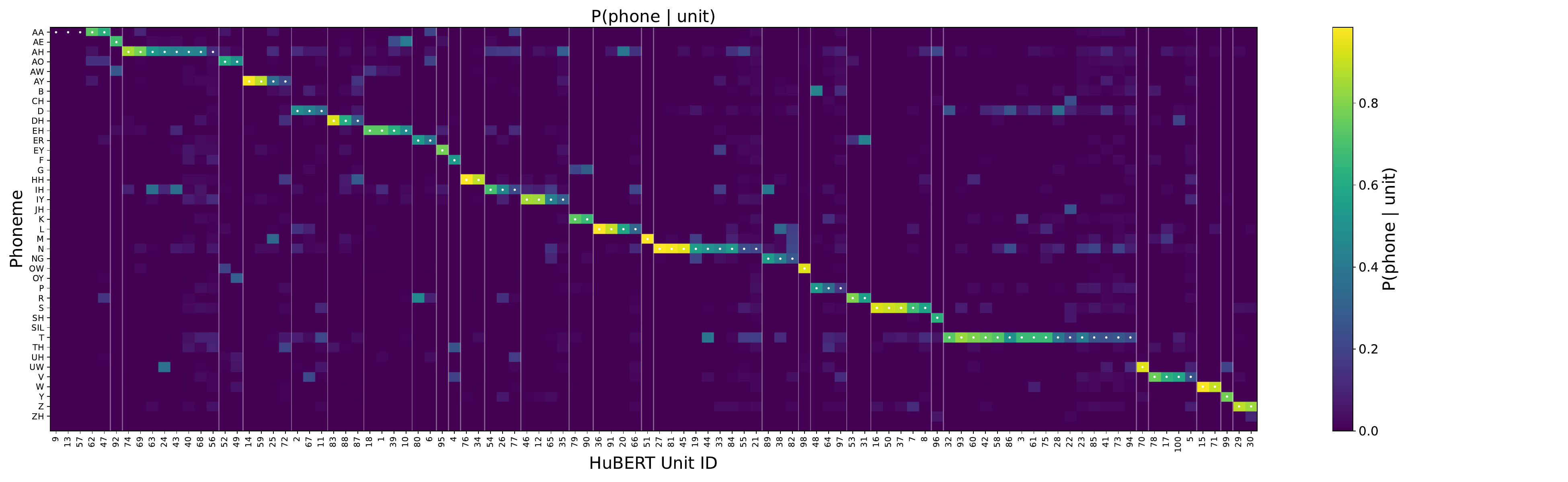}
  \caption{For each \textsc{HuBERT} unit, the phoneme with the highest conditional
probability (i.e., \(\texttt{argmax}_{\textsc{phoneme}} P(\textsc{phoneme}\mid \text{\em u})\)) is shown.
Multiple \textsc{HuBERT} units may map to the same most-probable phoneme \(\text{\em p}\).
In such cases, for each phoneme \(\text{\em p}\), we order the corresponding units from left to right
by increasing entropy of \(P(\cdot\mid \text{\em u})\).
}
  \label{fig:pGu}
\end{figure*}

We further probe the structure of \(P(\textsc{phoneme}\mid \texttt{dis}(\mathcal{H})_{\textsc{HuBERT}})\).
In figure~\ref{fig:pGu}, we visualize, for each \textsc{HuBERT} unit, the phoneme with the highest conditional
probability (i.e., \(\texttt{argmax}_{\textsc{phoneme}} P(\textsc{phoneme}\mid \text{\em u})\)).
Multiple \textsc{HuBERT} units may map to the same most-probable phoneme \(\text{\em p}\).
In such cases, for each phoneme \(\text{\em p}\), we order the corresponding units by increasing entropy of
\(P(\cdot\mid \text{\em u})\); lower entropy indicates a sharper, more confident association between the unit and
phoneme \(\text{\em p}\).
We find that alveolar consonants (e.g., \texttt{T}, \texttt{S}, \texttt{N}) and  \texttt{A} and \texttt{I} based vowels (e.g., \texttt{AA}, \texttt{AY}, \texttt{AH}, \texttt{IY}, \texttt{IH}) have many \textsc{HuBERT} units mapping to them.

This is consistent with prior analyses showing that these phones are among the most frequently occurring in conversational English and that their phonetic realizations have different manifestations depending on coarticulatory context \citep{mines1978frequency}.
These observations suggest that our lookup dictionary
\(P(\textsc{phoneme}\mid \texttt{dis}(\mathcal{H})_{\textsc{HuBERT}})\)
captures meaningful phonetic structure and is grounded in known articulatory regularities of speech.

\section{Detailed literature review}
\label{apd:Lit}

Here, we review prior work on speech neural and neuromuscular interfaces and contextualize our results relative to state-of-the-art methods.
A substantial body of research \cite{jou2006towards, kapur2020non, meltzner2018development, toth09_interspeech, 8114359, 8578038, littlejohn2025streaming} has laid the groundwork for EMG-based speech interfaces.
Among the earliest studies, \citet{jou2006towards} demonstrate EMG-to-speech conversion on a small corpus of 50 sentences.
\citet{kapur2020non} use a corpus of 15 sentences and, rather than performing phoneme-level decoding, formulate the task as a 15-way classification problem.
\citet{meltzner2018development} study EMG-to-text recognition for isolated words, phrases drawn from a $\sim$200-word vocabulary, and continuous sentences using a custom grammar-based recognition model over a set of 1200 scripted phrases.
\citet{toth09_interspeech} present EMG-to-speech conversion on a corpus of 500 sentences.
\citet{8114359} demonstrate EMG-to-speech conversion using up to two hours of data and 2000 utterances.

Overall, these studies rely on private datasets and task-specific pipelines, and they typically evaluate on small, constrained corpora.
In addition, the works do not release full implementations (e.g., code repositories) or sufficient methodological details to enable direct reproducibility.
As a result, it is difficult to directly compare performance across systems, and all the above results do not establish generalization to open-vocabulary English settings. 

A reproducible benchmark for open-vocabulary EMG-to-speech conversion was introduced by \citet{gaddy2020digital, gaddy2021improved}. However, these works rely on time-aligned EMG-audio pairs for training. Building on \citet{gaddy2021improved}, \citet{benster2024cross} propose an approach that leverages an audio-only corpus in addition to paired EMG-audio data. While effective in the benchmark setting, such methods can be difficult to deploy in clinical scenarios where parallel EMG-audio recordings may be unavailable or unreliable. On the large-vocabulary corpus, \citet{gaddy2020digital} report a word error rate (WER) of 68\%, and \citet{gaddy2021improved} reduce this to 42\%. \citet{benster2024cross} report WER $< 10\%$ on the \citet{gaddy2020digital} dataset by using a large language model (LLM) to post-correct the intermediate EMG-to-phoneme output. However, their system no longer supports streaming synthesis, and the evaluation does not fully characterize potential information leakage through the LLM (e.g., memorization or exposure to overlapping text distributions). Consequently, their results are not directly comparable to strictly streaming EMG-to-speech systems. \citet{littlejohn2025streaming} report a WER of 74\% on the \citet{gaddy2020digital} dataset using a CNN+RNN transducer model; however, their train-test splits and implementation details are not publicly available, which prevents direct comparison.
In our setting, we address a harder learning problem by not assuming time-aligned EMG-audio pairs during training, and we report a WER of 62\% on an open-vocabulary corpus. We emphasize that these WER values should not be compared one-to-one across studies, since the data collection setup, training targets and alignment assumptions, problem formulation, and evaluation methodology differ substantially.
We report these results to provide context relative to prior EMG-based speech interfaces.

To address these limitations, we build on widely available pretrained speech models and vocoders, but adapt them to the EMG setting through principled, articulatorily motivated design choices. We characterize the structure of orofacial EMG signals and introduce methods that are straightforward to implement yet carefully tailored for this problem, yielding substantial gains. For example, we represent EMG signals using covariance matrices and propose phoneme-guided decoding of speech units (section~\ref{sec:results_gen}), which exploits phonetic structure to improve the fidelity of synthesized speech. Our encoder design and phoneme-guided decoding are grounded in articulatory mechanisms and established regularities of speech production.

The WERs reported in this work fall broadly within the range observed for invasive speech neural interfaces.
For example, \citet{wairagkar2025instantaneous} report a median WER of 43\% using 256 intracortical electrodes, trained on approximately 8300 cued sentences spanning about 38 hours of data; the large number of hours for a comparable number of sentences reflects the substantially slower speaking rate in that study relative to ours.
Similarly, \citet{metzger2023high} report a WER of 54\% using 253 electrodes with a 1024-word vocabulary.
In our setting, we achieve a WER of 62\% on a large-vocabulary corpus.
These error rates should be interpreted in the context of neural speech interfaces, where the sensing modality, signal-to-noise ratio, data size, and evaluation setup differ substantially from conventional audio ASR (automatic speech recognition) benchmarks.

\section{Additional technical details}
Here, we provide a detailed description of the model architectures, training procedures, participant instructions, and ablation studies used throughout this work, with the goal of making our methodology clear, reproducible, and easy to interpret.
\label{apd:ablation}
\subsection{Effect of training data size}
We study how training-set size affects the unit error rate (UER) when decoding
$\texttt{dis}(\mathcal{H})_{\textsc{HuBERT}}$.
Using $\textsc{Data}_{\textsc{ general}}$, we train the model with between
2000 and 8000 sentences, while keeping the validation and test splits fixed
as in section~\ref{sec:Data}.
Over this range, we observe an approximately linear improvement in UER with
increasing training data.
This differs from the power-law behavior commonly reported for large-scale
from-scratch training \citep{kaplan2020scaling}.
Although we cannot draw strong conclusions from a single-participant study,
these results suggest that, when leveraging frozen pretrained speech
representations, performance may remain data-limited and continue to benefit
predictably from additional EMG training data, yielding lower UER (and
consequently WER) as the dataset size scales.

\begin{figure}[ht] 
    \centering 
    \includegraphics[width=0.48\textwidth]{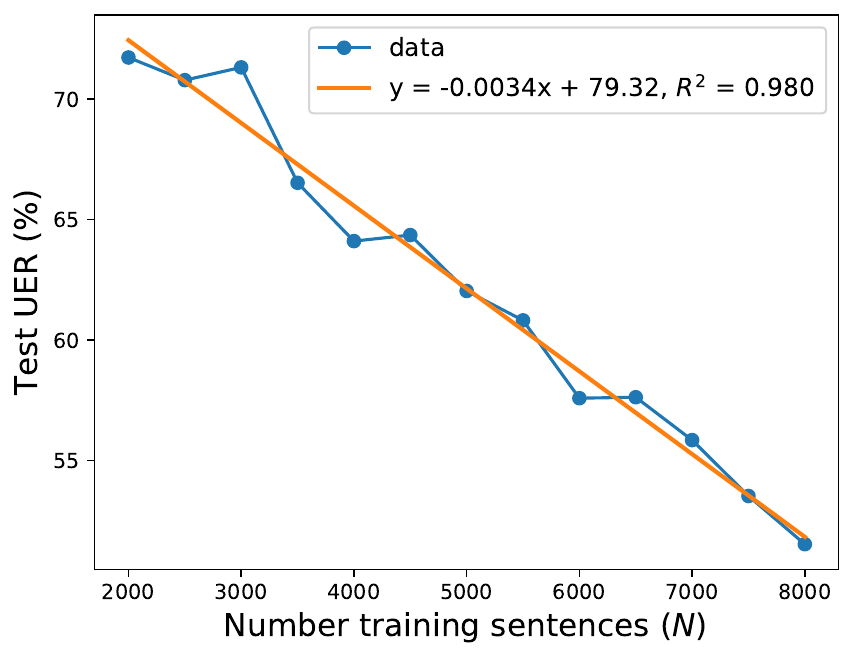} 
    \caption{Test UER as a function of the number of training sentences. Over the evaluated
range (2000-8000 sentences), UER decreases approximately linearly with
increasing training data, indicating consistent gains from additional EMG data. UER is computed by decoding $\texttt{dis}(\mathcal{H})_{\textsc{HuBERT}}$
from a model trained with $\texttt{vec}(\mathcal{E})$ as input; during training,
we use phoneme-guided decoding.
}
    \label{fig:werDataSize} 
\end{figure}

\subsection{Losses}
\begin{figure}[ht]
    \centering
    \includegraphics[width=0.48\textwidth]{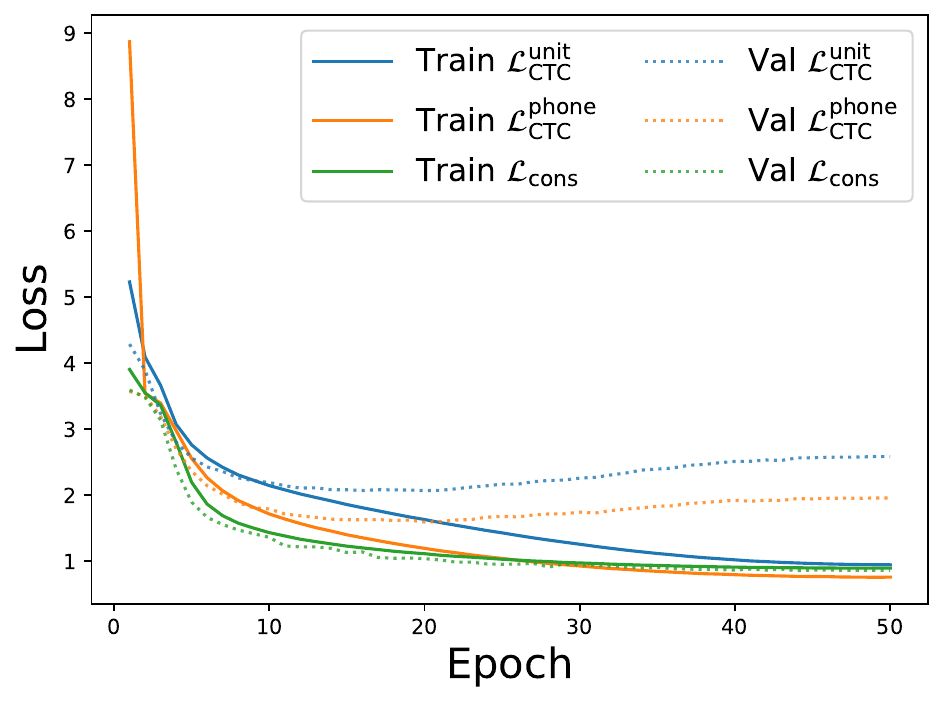}
    \caption{Training and validation losses versus epoch. The causal TDS convolutional model in section~\ref{sec:seq2seq} is trained for 50 epochs with $\texttt{vec}(\mathcal{E})$ as input, using $\mathcal{L}_{\mathrm{CTC}}^{\mathrm{unit}}$, $\mathcal{L}_{\mathrm{CTC}}^{\mathrm{phone}}$, and $\mathcal{L}_{\mathrm{cons}}$.}
    \label{fig:losses}
\end{figure}

As shown in figure~\ref{fig:losses}, all loss terms decrease smoothly during the
early stages of training, indicating stable optimization.
Validation losses begin to increase after approximately 25 epochs, suggesting
the onset of overfitting.
The consistency loss $\mathcal{L}_{\mathrm{cons}}$ and the phoneme-level CTC loss
$\mathcal{L}_{\mathrm{CTC}}^{\mathrm{phone}}$ decrease more rapidly than the
unit-level loss $\mathcal{L}_{\mathrm{CTC}}^{\mathrm{unit}}$, consistent with
their role as auxiliary objectives that regularize training and encourage
better alignment for unit prediction.
Together with figure~\ref{fig:werDataSize}, which shows that UER decreases
approximately linearly with training set size over the evaluated range and has
not yet saturated, and the overfitting observed in
figure~\ref{fig:losses}, these results suggest that the model remains
data-limited and can continue to benefit from additional training data.

\subsection{Transcriptions}
\label{apd:human}
Human evaluation of synthesized speech is commonly used to assess the
intelligibility and overall quality of generated audio \citep{wolters2010evaluating}.
For both $\textsc{Data}_{\textsc{ general}}$ and $\textsc{Data}_{\textsc{ ALS}}$,
we ask human raters to listen to the synthesized audio and transcribe each
utterance in English.
Raters are not restricted to a predefined vocabulary and may write any English
words for both corpora (even though $\textsc{Data}_{\textsc{ ALS}}$ contains only
about 300 unique words, raters are not informed of this constraint).
In this sense, our evaluation targets open-vocabulary recognition and is less
constrained than evaluations such as \citet{metzger2023high} and
\citet{littlejohn2025streaming}, which use fixed vocabularies of 1{,}024 words.

\begin{figure}[ht]
    \centering
    \includegraphics[width=0.48\textwidth]{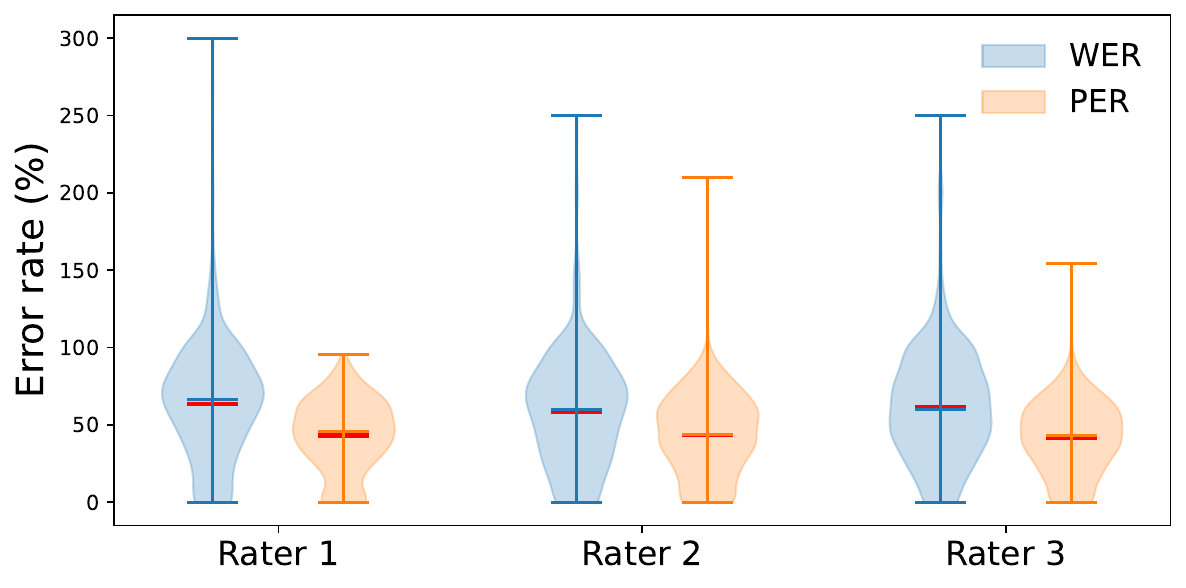}
    \caption{Distributions of WER and PER across three human transcribers for 460 sentences. Means are shown in red.}
    \label{fig:werPerViolin}
\end{figure}

In figure~\ref{fig:werPerViolin}, we summarize the distribution of WER and PER
across three human transcribers for all evaluated sentences (460 sentences in
total across $\textsc{Data}_{\textsc{ general}}$ and $\textsc{Data}_{\textsc{ ALS}}$).
The raters exhibit similar central tendency and spread, indicating strong
agreement.
To compute PER, we phonemize each rater's transcription and compare it against
the ground-truth phonemized sentence.

Across raters, the mean PER is 42.79\%, which is lower than the mean WER of
61.02\%.
This suggests that, even when transcribed words are incorrect, the errors are
often phonetically plausible.

In addition, we evaluate perceptual quality using mean opinion score (MOS). Specifically, we randomly sample 25 sentences from the 400-sentence test set of $\textsc{Data}_{\textsc{ general}}$ and 10 sentences from the 60-sentence test set of $\textsc{Data}_{\textsc{ ALS}}$, yielding 35 sentences in total. We then ask 10 native English speakers to rate the quality of the synthesized audio on a 1-to-5 MOS scale, resulting in 350 individual ratings. We restrict MOS evaluation to this subset because three human listeners had already listened to and transcribed all 460 test sentences for the WER/PER evaluation shown in figure~\ref{fig:werPerViolin}, and collecting MOS ratings for the full test set from a larger pool of raters was not practically feasible.

\begin{figure}[ht]
    \centering
    \includegraphics[width=0.48\textwidth]{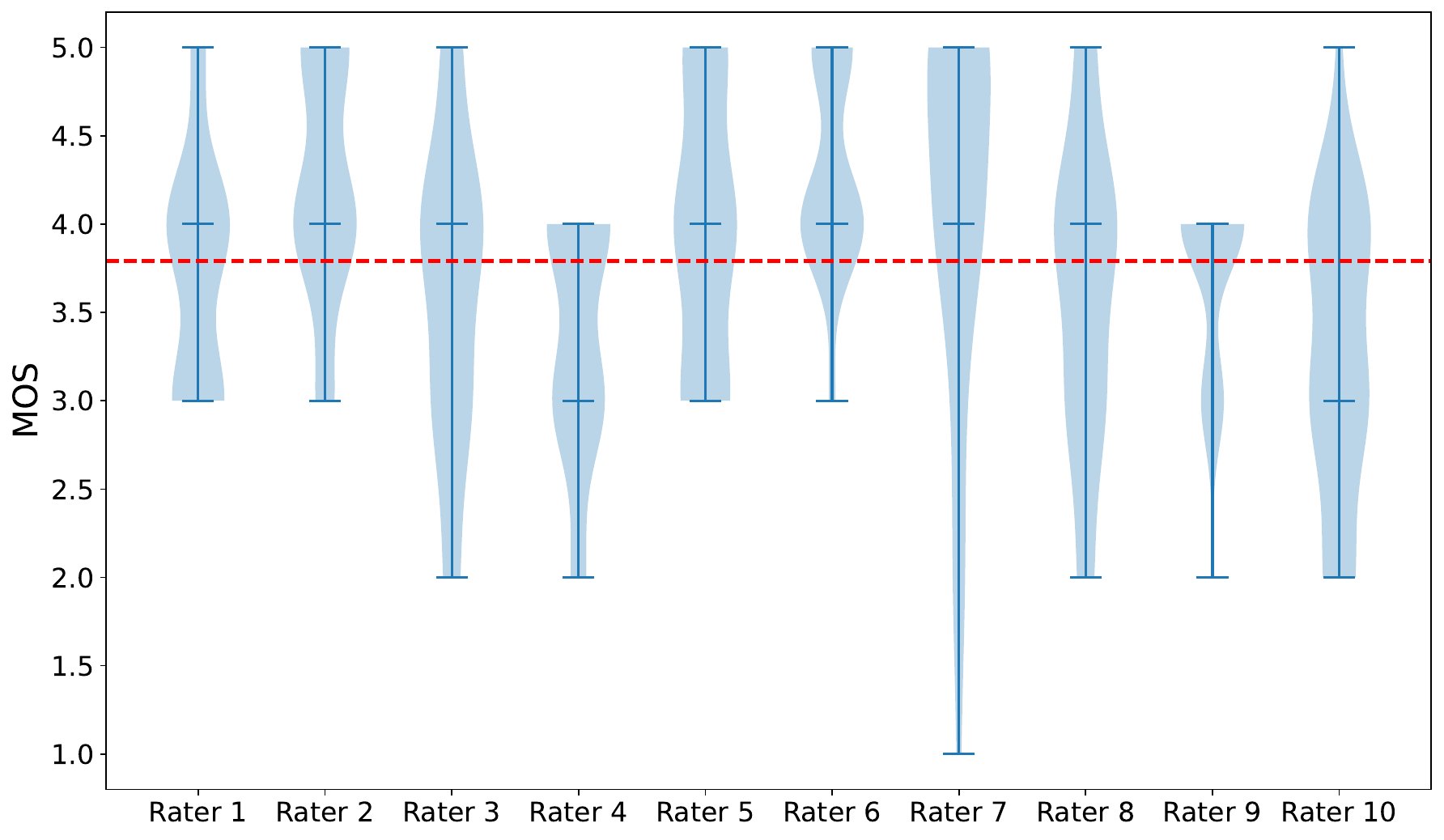}
    \caption{Distribution of mean opinion score (MOS) ratings across 10 raters for 35 synthesized sentences. Mean MOS across raters is shown in red.}
    \label{fig:mos}
\end{figure}

Figure~\ref{fig:mos} shows the distribution of MOS ratings across the 10 raters. The mean MOS across all ratings is 3.79. This value should be interpreted in the context of our generation setting. In conventional text-to-speech or voice conversion evaluations, the input typically contains clean linguistic content, and MOS primarily reflects the naturalness and perceptual quality of the generated speech. In our setting, by contrast, speech is synthesized directly from EMG, where the underlying linguistic representation is itself noisy and may already be corrupted prior to waveform generation. In addition, we use a pretrained vocoder \citep{lakhotia-etal-2021-generative} that was trained on natural speech rather than on linguistically corrupted intermediate representations. Consequently, the synthesized outputs can remain acoustically speech-like and perceptually plausible even when the recovered linguistic content is inaccurate. This helps explain, at least in part, why MOS remains moderately high despite high WER: in our setting, MOS reflects the perceptual plausibility of the synthesized waveform, whereas WER reflects the fidelity of the recovered linguistic content.

We open-source all 460 synthesized sentences from the test set, together with the ground-truth audio for all 9,660 sentences from the healthy participant. Ground-truth audio for the participant with ALS is unavailable because she articulated the sentences {\em silently}. These audio samples are available on the \textsc{Project Page}. To provide a reference point for future work, we transcribed the healthy participant’s ground-truth audio using \textsc{Whisper} \citep{radford2023robust}, obtaining a WER of 14.25\% on the training-validation split and 12.66\% on the test set. These values should be treated as baseline references for future improvements.

\subsection{emg2text}
Furthermore, we train the TDS convolutional model to decode phonemes on all 460
sentences (following the procedure in section~\ref{sec:seq2seq}) and then map
the predicted phoneme sequences to words using a weighted finite-state
transducer (WFST) decoder\footnote{We use the WFST decoding implementation provided by \textsc{icefall}
\(\left(\raisebox{-0.2ex}{\includegraphics[height=3ex]{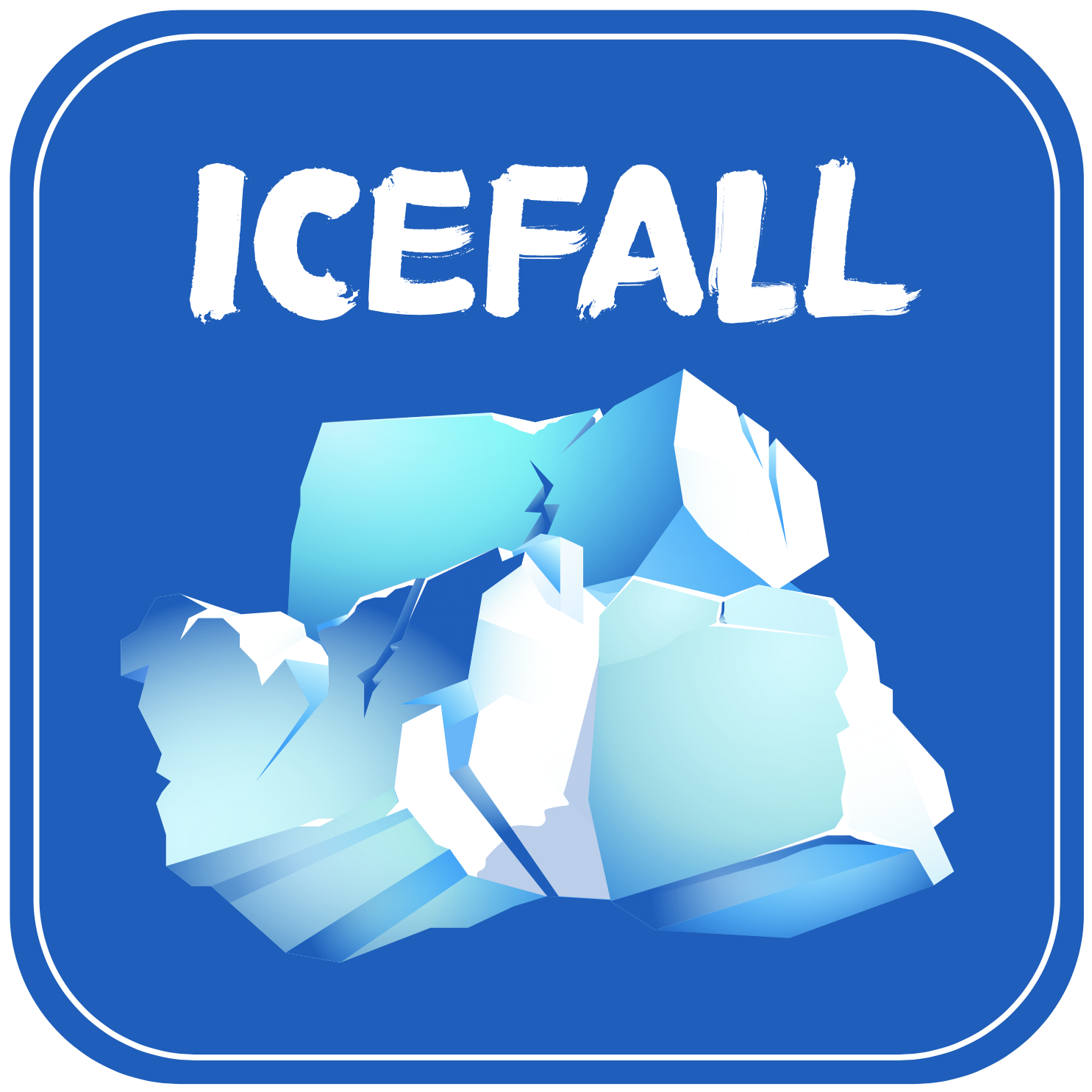}}
\href{https://github.com/k2-fsa/icefall}{\texttt{ github.com/k2-fsa/icefall}}\right).\)}.

For phoneme-to-word decoding, we use the LibriSpeech-100 transcripts
\citep{7178964}, which contain roughly 38000 sentences and 35000 unique
words. From these transcripts, we build a pronunciation lexicon WFST, {\em L},
that maps phoneme sequences to words. We also train a 4-gram language model
with KenLM \citep{heafield-2011-kenlm} and convert it into a grammar WFST, {\em G}.
Finally, we construct the CTC topology WFST, {\em H}, which encodes the allowable
label sequences under the CTC criterion.

We compose these components into a decoding graph \citep{35080},
$\text{\em HLG} = \text{\em H} \circ \text{\em L} \circ \text{\em G}$,
which integrates the CTC constraints ({\em H}), lexicon mapping ({\em L}), and language
model probabilities ({\em G}). At inference time, we perform beam search over
{\em HLG} with beam width 50 and compute WER as the normalized Levenshtein
distance between the reference and decoded word sequences. This procedure
yields a WER of 51.17\% and a PER of 38.19\%. The language model is trained only
on LibriSpeech-100 transcripts; sentences from our train-validation-test splits
are not included. We summarize this approach, denoted \textsc{emg2text} with direct \textsc{emg2speech} in table~\ref{tab:textVspeech}.

\begin{table}[h!]
\centering
\caption{PER and WER for \textsc{emg2text} and \textsc{emg2speech}.}
\setlength{\tabcolsep}{4pt}
\renewcommand{\arraystretch}{1.05}
\begin{tabular}{lcc}
\hline
\textsc{\footnotesize Training} &
\textsc{\footnotesize {PER} (\footnotesize$\%\downarrow$)} &
\textsc{\footnotesize {WER} (\footnotesize$\%\downarrow$)}\\
\textsc{\footnotesize method} & & \\
\hline
\textsc{emg2speech} & 42.79 & 61.02 \\
\textsc{emg2text} & 38.19 & 51.17 \\
\hline
\end{tabular}
\label{tab:textVspeech}
\end{table}

Overall, \textsc{emg2text} achieves lower PER and WER than direct
\textsc{emg2speech}.
However, direct EMG-to-speech generation remains important for neural
prostheses because it can enable a more fluid, natural interaction (e.g.,
without requiring an explicit intermediate text interface).
We therefore view improving direct EMG-to-speech as an important direction for
future work.\\

\subsection{Model architecture and implementation details}

\textbf{Input representation.} Let $\mathbf{x} \in \mathbb{R}^{\text{\em T} \times \text{\em N} \times \texttt{dim}}$ denote the input, where {\em N} is the batch size, {\em T} is the number of time steps, and \texttt{dim} denotes the dimensionality of the input EMG representation, namely $\texttt{vec}(\mathcal{E})$, $\mathbb{D}(\mathcal{E})$, or $\texttt{vec}(\mathcal{B})$.

\noindent \textbf{Channel normalization.} Before the multi-layer perceptron (MLP), we apply channel-wise batch normalization to the input. For $\texttt{vec}(\mathcal{E})$ and $\texttt{vec}(\mathcal{B})$, we reshape $\mathbf{x}$ to $\mathbb{R}^{\text{\em T} \times \text{\em N} \times \mathcal{V} \times \mathcal{V}}$ and $\mathbb{R}^{\text{\em T} \times \text{\em N} \times \mathcal{V} \times \text{\em B}}$, respectively, and apply 2D batch normalization with $\mathcal{V}$ as the channel dimension. For $\mathbb{D}(\mathcal{E})$, we apply 1D batch normalization directly to $\mathbf{x} \in \mathbb{R}^{\text{\em T} \times \text{\em N} \times \mathcal{V}}$. In all cases, normalization statistics are computed independently for each channel $\mathcal{V}$ across the batch and temporal dimensions, and additionally across the final feature dimension when present. In our case, $\mathcal{V}$ = 31 and {\em B} = 31.

\noindent \textbf{Spatially robust MLP.} After channel normalization, we apply an MLP frontend designed to improve robustness to spatial variability across electrodes. Specifically, we construct an ensemble of views by circularly shifting the input along the channel dimension with offsets $o \in \mathcal{O}$, where $\mathcal{O} = \{-1,0,1\}$. For matrix-valued inputs, i.e., $\texttt{vec}(\mathcal{E})$ or $\texttt{vec}(\mathcal{B})$, of shape $\mathbb{R}^{\text{\em T} \times \text{\em N} \times \mathcal{V} \times \mathcal{F}}$, where $\mathcal{F}=\mathcal{V}$ for $\texttt{vec}(\mathcal{E})$ and $\mathcal{F}=\text{\em B}$ for $\texttt{vec}(\mathcal{B})$, each shifted view is flattened across the last two dimensions and passed through a shared MLP. For $\mathbb{D}(\mathcal{E}) \in \mathbb{R}^{\text{\em T} \times \text{\em N} \times \mathcal{V}}$, we analogously shift along the channel dimension and apply the same shared MLP directly. The resulting embeddings are aggregated across shifts using mean pooling:
\begin{align*}
\mathbf{h}
&=
\frac{1}{|\mathcal{O}|}
\sum_{o \in \mathcal{O}}
\mathrm{MLP}\!\left(\phi\!\left(\mathrm{shift}(\mathbf{x}, o)\right)\right), \\
&
\mathcal{O} =
\{-1,0,1\},
\end{align*}
where $\phi(\cdot)$ denotes flattening for matrix-valued inputs and the identity map for $\mathbb{D}(\mathcal{E})$. This yields $\mathbf{h} \in \mathbb{R}^{\text{\em T} \times \text{\em N} \times \text{\em H}}$, where {\em H} is the output dimensionality of the final MLP layer. In our case, {\em H} = 384.

\noindent \textbf{TDS convolutional encoder.} The output of the spatially robust MLP, denoted by $\mathbf{h} \in \mathbb{R}^{\text{\em T} \times \text{\em N} \times \text{\em H}}$, is passed to a temporal encoder based on the time-depth separable (TDS) architecture of \citet{hannun2019sequence}. For a given TDS block, let $\mathbf{u} \in \mathbb{R}^{\text{\em T} \times \text{\em N} \times \text{\em H}}$ denote the block input. We reshape $\mathbf{u}$ to $\mathbb{R}^{\text{\em N} \times \text{\em K} \times \text{\em W} \times \text{\em T}}$, where {\em H = KW}. A causal convolution with kernel tensor $\Theta \in \mathbb{R}^{\text{\em K} \times \text{\em K} \times 1 \times \text{\em k}}$ and kernel size $1 \times \text{\em k}$ is then applied along the temporal dimension, with {\em k} = 14 and replicate padding of length $\text{\em k}-1$ on the left:
\begin{align*}
\tilde{z}_{n,k,w,t}
&=
\sum_{i=0}^{k-1}
\sum_{k'=1}^{K}
\Theta_{k,k',0,i}\,
\tilde{u}_{n,k',w,t-i}, \\
\Theta
&\in
\mathbb{R}^{\text{\em K} \times \text{\em K} \times 1 \times \text{\em k}}.
\end{align*}
The convolution output is passed through a ReLU nonlinearity, reshaped back to $\mathbb{R}^{\text{\em T} \times \text{\em N} \times \text{\em H}}$, added to the block input through a residual connection, and normalized with LayerNorm:
\begin{align*}
\mathbf{z}_{\mathrm{conv}}
&=
\mathrm{LayerNorm}\!\left(
\mathrm{ReLU}(\mathrm{Conv}(\mathbf{u}))
+
\mathbf{u}
\right), \\
&\mathbf{z}_{\mathrm{conv}}
\in
\mathbb{R}^{\text{\em T} \times \text{\em N} \times \text{\em H}}.
\end{align*}
The subsequent fully connected block applies two linear layers with a ReLU nonlinearity in between, followed by a residual connection and LayerNorm:
\begin{align*}
\mathbf{z}_{\mathrm{fc}}
&=
\mathrm{LayerNorm}\!\left(
\mathrm{FC}_2\!\left(
\mathrm{ReLU}\!\left(
\mathrm{FC}_1(\mathbf{z}_{\mathrm{conv}})
\right)\right)
+
\mathbf{z}_{\mathrm{conv}}
\right), \\
&\mathbf{z}_{\mathrm{fc}}
\in
\mathbb{R}^{\text{\em T} \times \text{\em N} \times \text{\em H}}.
\end{align*}
We stack four such TDS blocks with channel configuration [24, 24, 24, 24]. Since the MLP output dimensionality is fixed to {\em H} = 384, each block uses width {\em W} = 384/24 = 16. All convolutions are causal.

\noindent \textbf{Dual output heads.} Following the encoder, a shared bottleneck maps {\em H}-dimensional representations to a 512-dimensional space. Two independent linear heads then produce log-softmax outputs: a unit head \texttt{Linear}(512, 101) over 100 \textsc{HuBERT} units plus a CTC blank symbol (index 100), and a phone head \texttt{Linear}(512, 41) over 40 phonemes plus a CTC blank symbol (index 40).

\noindent \textbf{Optimization.} We use AdamW ($\beta_1=0.9$, $\beta_2=0.98$, weight decay = $10^{-4}$, learning rate $= 3 \times 10^{-4}$) with a linear warm-up over 5 epochs, starting from $0.1 \times$ the base learning rate, followed by cosine annealing to a minimum of $10^{-6}$ over the remaining 45 epochs. Models are trained for 50 epochs with per-epoch temporal jitter resampling of the training set.

\end{document}